\documentclass[11pt]{article}
\usepackage{amsmath,amssymb}
\usepackage{xcolor}
\usepackage{framed}

\usepackage{graphicx}

\usepackage{amsmath,latexsym}

\begin{document}

\setcounter{page}{1}

\pagestyle{plain}

\begin{center}
\Large{\bf Quantum Gravity Effects on the Tachyon Inflation from Thermodynamic Perspective}\\
\small \vspace{1cm} {\bf M. Bitaj \footnote{mozhdeh.bitaj@gmail.com}},\quad{\bf N. Rashidi
\footnote{n.rashidi@umz.ac.ir (Corresponding Author)}\,, \quad \bf
K. Nozari \footnote{knozari@umz.ac.ir }\quad and \quad
\bf M. Roushan \footnote{m.roushan@umz.ac.ir}}  \\
\vspace{0.5cm} Department of Theoretical Physics, Faculty of
Science,
University of Mazandaran,\\
P. O. Box 47416-95447, Babolsar, IRAN
\end{center}

\begin{abstract}
By considering the Friedmann equations emerging from the
entropy-area law of black hole thermodynamics in the context of the
generalized uncertainty principle, we study tachyon inflation in the
early universe. The presence of a minimal length modifies the
Friedmann equations and hence the slow-roll and perturbation
parameters in the tachyon model. These modifications, though small,
affect the viability of the tachyon inflation in confrontation with
observational data. We compare the numerical results of the model
with Planck2018 TT, TE, EE +lowE+lensing+BAO+BK14(18) data and
Planck2018 TT, TE,EE +lowE+lensing+BK14(18) +BAO+LIGO $\&$ Virgo2016
data at $68\%$ and $95\%$ CL. We show that while the tachyon
inflation with power-law, inverse power-law and inverse exponential
potentials is not observationally viable in comparison with the
$1\sigma$ and $2\sigma$ confidence levels of the new joint data, in
the presence of the minimal length the model becomes observationally viable.
\end{abstract}
\newpage

\section{Introduction}

{All approaches to quantum gravity proposal predict the existence of a fundamental length scale of the order of the Planck (or fundamental string) length, below of which no distinction/resolution of the spacetime points is possible~\cite{Ama89}. In fact, black hole physics~\cite{Cas15} together with all phenomenological approaches to QG proposal including String Theory~\cite{Ama89,Gro88,Yo89,Ko90}, Loop Quantum Gravity~\cite{Rov95,Ash97,Ash98,Mo09}, Asymptotically Safe Quantum Gravity~\cite{Lau05,Re06,Per10,Eic19}, Noncommutative Geometry~\cite{Sei99,Con99}, Polymer Quantization of spacetime~\cite{Ash03} and some Gedanken experiments (thought experiments) proposed to unify Quantum Mechanics and General Relativity~\cite{Hos13},  all predict the existence of a minimal fundamental measurable length of the order of the Planck length. To incorporate this minimal length in ordinary quantum mechanics, one modifies the Heisenberg Uncertainty Principle to derive the so-called Generalized (Gravitational) Uncertainty Principle (GUP). A GUP setup includes the gravitationally induced position uncertainty associated with the minimal fundamental length scale. On the other hand, the Tachyon inflation has its origin in string theory tachyon condensation and as is known~\cite{Kof02}, cosmic inflation in these theories occurs at super-Planckian densities; the era of which QG effects are important in essence. If this is the case and as far as we know it is definitely the case, quantum gravity effects encoded in the existence of a minimal length/maximal energy(momentum) have to be incorporated in a Tachyon inflation. In fact, the minimal length/maximal energy as a ultra-violet natural cutoff encoded in the generalized uncertainty principle governs generation of seeds and evolution of the cosmological perturbations. This feature technically is addressed by imposing the existence of the minimal length on the wavelength of acoustic perturbations or equivalently on wave numbers (or equivalently wave’s momentum). This is the main motivation for incorporation of the generalized uncertainty principle in this Tachyon inflation scenario.}

{It is worth noting that some subtleties in transition to relativistic field theory, such as self-adjointness of the position and momentum operators can be addressed via von Neumann theorem through maximally localized states or quasi-position/momentum states in dense domains of these operators in the generalized Bargmann-Fock space (See~\cite{Kem95} and Appendix C of~\cite{No12} for details). About the self-adjointness of the time operator, formally a similar argument is possible in essence, but not in a maximally localized states (or quasi-position/momentum states) in position/momentum spaces, but in a time/energy formal space. Also, doubly special relativity provides a maximal momentum~\cite{Ame02,Cor05} that regularizes the free particle’s energy not to be divergent anymore and position (momentum)  is actually an operator in the dense domain of $\{x,x^{2}\}$ ($\{p,p^{2}\}$) in maximally localized states or quasi-position (quasi-momentum)  representation in Bargmann-Fock space. The details of such arguments can be seen in Refs.~\cite{Kem95} and~\cite{No12}. }

In the study of the relativistic cosmology, the Friedmann equations
are two important equations governing the dynamics of the universe.
These equations, which are obtained from the Einstein field
equations, have great importance to describe the expansion history
of the universe. Interestingly, it has been shown that the Einstein
equations can be obtained from the thermodynamic viewpoint by using
the Clausius relation between the temperature ($T$), entropy ($S$),
and heat ($Q$)~\cite{Jac95}. This idea was based on the connection
between entropy and horizon area in black hole physics and demanding
that through each spacetime point the Clausius relation $dQ=T\,dS$
holds for all the local Rindler causal horizons. In this context,
the Einstein field equations are claimed to be an equation of state
for the considered spacetime (see~\cite{Jac95} for more details). On
the other hand, Verlinde~\cite{Ver00} has shown that the Friedmann
equation governing the dynamics of the FRW spacetime can be
rewritten in the same form as the entropy formula which governs the
thermodynamics of radiation component in the universe. See
also~\cite{Noj01,Cai03,Noj02,Cve02} for more literature on this
issue. Overall, it seems that there are some relations between
thermodynamics and Einstein equations and this is an interesting
subject to be studied in more relevant backgrounds. In this regard,
the authors of~\cite{Cai05} have obtained ($n+1$)-dimensional
Friedmann equation for the FRW universe, based on the thermodynamic
perspective. Their approach was based on the following relations for
the temperature $T$ and entropy $S$ of the apparent horizon
\begin{equation}
\label{eq1} T=\frac{1}{2\pi \tilde{r}_{A}}\,,
\end{equation}
\begin{equation}
\label{eq2} S=\frac{A}{4G}\,,
\end{equation}
where $\tilde{r}_{A}$ is the apparent horizon radius of the
universe, $G$ is the Newton constant and $A$ is the apparent horizon
area. They have started their work by considering the following
($n+1$)-dimensional FRW metric
\begin{equation}
\label{eq3}
ds^2=-dt^2+a^{2}(t)\left(\frac{dr^2}{1-k\,r^2}+r^2\,d\Omega_{n-1}^{2}\right)\,,
\end{equation}
where $k=0,\pm 1$ shows the spatial curvature constant and
$d\Omega_{n-1}^{2}$ is the line element of an $(n-1)$-dimensional
unit sphere. By rewriting the line element (\ref{eq3}) as
\begin{equation}
\label{eq4}
ds^2=h_{ij}\,dx^{i}\,dx^{j}+\tilde{r}^2\,d\Omega_{n-1}^{2}\,,
\end{equation}
where $h_{ij}$ is a two-dimensional metric as
$(-1,\,a^{2}/(1-kr^2))$, $\tilde{r}= a(t)\,r$ and $x^{i}=(t,r)$,
they have obtained the apparent horizon $\tilde{r}_{A}$ in terms of
the Hubble parameter $H$. Then, by using the first law of
thermodynamics and the relation between the apparent horizon
temperature and the surface gravity, they obtained the Friedmann
equation for an ($n+1)$-dimensional universe as
\begin{equation}
\label{eq5} H^2=\frac{16\,\pi\,G}{n(n-1)}\rho-\frac{k}{a^2}\,,
\end{equation}
where $\rho$ is the energy density of the components inside the
apparent horizon. See Ref.~\cite{Cai05} for more details on
obtaining Friedmann equation (\ref{eq5}).

Another interesting subject is the connection between the
uncertainty principle and Hawking radiation in the black hole
physics context~\cite{Med04,Adl01,Cav03,Cav04,Maj11,Ali12a,Ali12b}.
Especially, the idea of the Generalized Uncertainty Principle (GUP),
proposed in black hole physics (and also string theory), has
attracted a lot of attention. There is a belief that there is a
minimal length implied by quantum gravitational considerations,
leading to the modified uncertainty principle as~\cite{Cav03}
\begin{equation}
\label{eq6} \Delta x_{i}\gtrsim \frac{\hbar}{\Delta
p_{i}}\left[1+\frac{\beta\, l_{Pl}^{2}}{\hbar^2}\left(\Delta
p_{i}\right)^{2}\right]\,,
\end{equation}
where $l_{Pl}$ is the Planck length and $\beta$ is a dimensionless
parameter whose value depends on the specific model. In the $\Delta
x_{i}\gg l_{Pl}$ limit (classical limit), the equation (\ref{eq6})
recovers the standard Heisenberg uncertainty relation as $\Delta
x_{i}\, \Delta p_{i}\geq \hbar \delta_{ij}$ or $\Delta x\, \Delta
p\geq \hbar$ in one space dimension. From now on, we adopt
$\hbar=1=c$. By considering equation (\ref{eq6}), and the argument
that the uncertainty principle $\Delta p \geq \frac{1}{\Delta x}$ is
equivalent to $\Delta E \geq \frac{1}{\Delta x}$~\cite{Cam04}, one can find
a lower bound on the energy of a test particle in the GUP case, assuming $E\sim\Delta E$.
Since the area of the black hole and correspondingly, its entropy is
related to the energy, the lower bound on energy gives a modified
expression for the entropy. This modification, via the first law of
thermodynamics, leads to a modified Friedmann equation~\cite{Awa14}.

One branch of cosmology, where the Friedmann equations play an
important role, is primordial cosmological inflation. In the
inflation era, the universe has expanded exponentially in a very
short time. A lot of interesting works on inflation have been done
which some of them predict gaussian distribution for dominant modes
of the perturbations while some others predict non-gaussian
distribution~\cite{Gut81,Lin82,Alb82,Lin90,Lid00a,Lid97,Rio02,Lyt09,Mal03}.
Among the various inflation models, the tachyon inflation is the one
with interesting results~\cite{Sen99,Sen02a,Sen02b,Gib02}. In the
tachyon inflation, the tachyon field (a non-canonical field
associated with the D-branes in string theory) drives the inflation
phase of the universe. In this model, the universe evolves smoothly
from the accelerating phase of the expansion to the non-relativistic
fluid-dominated phase. For some interesting works on tachyon
inflation see~\cite{Noj03,Bou16,Rez17}. Every inflation model, in
order to be considered viable, should be consistent with
observational data. To check the observational viability of an
inflation model, we can perform some numerical analysis on the
perturbation parameters such as the scalar and tensor spectral
indices ($n_{s}$) and ($n_{T}$), respectively and also the
tenor-to-scalar ratio ($r$). By comparing the results of the
numerical analysis with the newest observational data, we can test
the viability of the inflation model. {To this end,
we should adopt some types of potential. In Ref.~\cite{Maj04} the
authors have shown that the tachyon inflation can be considered as a
stringy implementation of the chaotic inflation. In this regard, it
seems interesting to consider chaotic potentials.} For instance, a
tachyon model with $\phi^{2}$ and $\phi^{4}$ potentials is not
consistent with observational data. {In
Ref.~\cite{Li14}, the tachyon inflation with both $\phi^2$ and
$\phi^4$ potentials has been considered. The authors have compared
this model with Planck2013 results~\cite{Ade14} and obtained the
observational viability of the model. According to their analysis,
for $\phi^2$ potential, the tachyon model has $n_{s}=0.975$
corresponding to $r=0.066$. Also, for $\phi^4$ potential, the
tachyon model has $n_{s}=0.972$ corresponding to $r=0.088$. Although
these values were consistent with Planck 2013 data, now they are out
of $1\sigma$ and $2\sigma$ confidence level of Planck2018 data. In
Ref.~\cite{Rez17}, the authors have considered several potentials
including $\phi^{-2}$, $\phi^{-4}$ and inverse exponential
potential. They have obtained some consistency between the tachyon
model with Planck2015~\cite{Ade16}. However, with Planck2018 model
these cases rule out.}

Given that in the early universe, the energy was very high, it seems
that the quantum gravitational effects should be important in the
physics of that era. {In fact, since existence of a minimal measurable length/ maximal measurable energy, is a phenomenological aspect of all existing approaches to QG proposal, it is natural to seek for its effects on a high energy phenomenon such as the cosmological inflation, the phenomenon that stringy effects are inevitable, at least in Tachyon condensation as the main underlying concept of this setup.  If stringy effects are important in high energy regime, then investigation of the effect of this high energy ingredient as a UV cutoff is essentially inevitable.} Therefore, it is interesting to consider the
quantum gravitational effects in the study of the primordial
inflation~\cite{Noj02,Noj03,Duf97}. In this paper, we include the
quantum gravitational effects in the early universe by considering
the modified Friedmann equations consisting minimal length effects.
In this way, the slow-roll parameters in the inflation model are
modified and therefore, the main perturbation parameters (scalar and
tensor spectral indices) are modified too. These modifications of
the perturbation parameters can change the status of the viability of the model in
confrontation with recent observational data. To test the viability of the
model, we consider both Planck2018 TT, TE,
EE+lowE+lensing+BAO+BK\textbf{14} data and Planck2018 TT, TE, EE
+lowE+lensing+BK\textbf{14}+BAO+LIGO and Virgo2016
data~\cite{pl18a,pl18b}. Also, we examine the observational
viability of this modified model in confrontation with Planck2018
TT, TE, EE+lowE+lensing+BAO+BK\textbf{18} data and Planck2018 TT,
TE, EE +lowE+lensing+BK\textbf{18}+BAO +LIGO and Virgo2016
data~\cite{Bi18a,Bi18b}.

With these preliminaries, this paper is organized as follows. In
section 2, we review the reconstruction of the Friedmann equations
from the first law of thermodynamics with general expression for the
entropy. In section 3, by considering the tachyon field in the
model, we present the modified Friedmann equations of the tachyon
model in the presence of a minimal length. In section 4, tachyon
inflation in the presence of the minimal length is formulated. In
section 5, we perform a numerical analysis on the model parameter
space and compare the results with several data sets. In this manner, we find the
viability of our model in some ranges of the deviation parameter. In
section 6, we present a summary of our work.

\section{Modified Friedmann Equations: Reconstruction from the First Law of Thermodynamics}

To reconstruct the modified Friedmann equations, we start with the
spatially flat ($3+1$) dimensional FRW background with the following
line element
\begin{eqnarray}
\label{eq7}
ds^2=h_{ij}\,dx^i\,dx^{j}+\tilde{r}^{2}\,d\Omega_{2}^{2}\,.
\end{eqnarray}
We can find the dynamic apparent horizon radius with this line
element. This horizon, a marginally trapped surface with the
vanishing expansion, is a sphere of radius $\tilde{r}_A$ that
satisfies the following equation~\cite{Cha10,Cha11}
\begin{eqnarray}
\label{eq8}
h^{ij}\,\partial_{i}\tilde{r}\,\partial_{j}\tilde{r}=0\,.
\end{eqnarray}
Solving this equation gives the following expression for the
apparent horizon in the flat FRW universe~\cite{Cai05}
\begin{eqnarray}
\label{eq9} \tilde{r}_{A}=a\,r=\frac{1}{H}\,,
\end{eqnarray}
where $H$ is the Hubble parameter defined as
$H=\frac{\dot{a}}{a}$. Note that, an over dot shows a cosmic time
derivative. Now, we consider a universe filled with perfect fluid
with the following energy-momentum tensor
\begin{eqnarray}
\label{eq10} T_{\mu\nu}=(\rho+p)\,u_{\mu}\,u_{\nu}+p\,g_{\mu\nu}\,.
\end{eqnarray}
In this definition, we have shown the $4$-velocity of the fluid with
$u_{\nu}$. Also, the energy density and pressure of the perfect
fluid are represented with $\rho$ and $p$, respectively. Satisfying
the conservation law for $T^{\mu\nu}$ gives the conservation
equation as
\begin{eqnarray}
\label{eq11} \dot{\rho}+3\,H\,(\rho+p)=0\,.
\end{eqnarray}
Following Refs.~\cite{Hay98,Bak00}, the work function or work
density in terms of the trace of the energy-momentum tensor is given
by
\begin{eqnarray}
\label{eq12} W=-\frac{1}{2}T^{ij}\,g_{ij}\,.
\end{eqnarray}
Note that in equation (\ref{eq12}), $T^{ij}$ is the projection of
the ($3+1$)-dimensional energy-momentum tensor $T^{\mu\nu}$ in the
normal direction on the 2-sphere~\cite{Cai05}. When we consider a
flat FRW universe filled with a perfect fluid, equation (\ref{eq12})
gives
\begin{eqnarray}
\label{eq13} W=\frac{1}{2}(\rho-p)\,.
\end{eqnarray}
This work function is important in treating the first law of
thermodynamics. The first law of thermodynamics is given by
\begin{eqnarray}
\label{eq14} dE=T\,dS+W\,dV\,,
\end{eqnarray}
where the parameter $E$ is the total energy inside the apparent
horizon or the Misner-Sharp energy and it is defined
as~\cite{Awa14,Cha11}
\begin{eqnarray}
\label{eq15} E=\rho\,V\,.
\end{eqnarray}
In the above definition, $V$ is the volume of a 3-dimensional sphere
with radius $\tilde{r}_{A}$, given by $V=\frac{4}{3}\pi
\tilde{r}_{A}^{3}$. From equation (\ref{eq15}) we get
\begin{eqnarray}
\label{eq16} dE=\rho\,dV+V\,d\rho\,.
\end{eqnarray}
Now, by using equations (\ref{eq11}), (\ref{eq15}), and (\ref{eq16})
we find
\begin{eqnarray}
\label{eq17}
dE=4\pi\tilde{r}_{A}^{2}\,\rho\,d\tilde{r}_{A}-4\pi\,\tilde{r}_{A}^{3}(\rho+p)\,H\,dt\,.
\end{eqnarray}
On the other hand, the term $W\,dV$ in equation (\ref{eq14}) is
given by
\begin{eqnarray}
\label{eq18}
W\,dV=2\pi\,\tilde{r}_{A}^{2}\,(\rho-p)\,d\tilde{r}_{A}\,.
\end{eqnarray}
Also, we need to rewrite the term $T\,dS$ in equation (\ref{eq14})
in terms of $\tilde{r}_{A}$. To this end, in Ref.~\cite{Awa14} the
authors have used the relationship between temperature and surface
gravity as
\begin{eqnarray}
\label{eq19} T=\frac{\kappa}{2\pi}\,.
\end{eqnarray}
The parameter $\kappa$ is the surface gravity defined as
\begin{eqnarray}
\label{eq20}
\kappa=\frac{1}{2\sqrt{-h}}\,\partial_{i}\bigg(\sqrt{-h}\,h^{ij}\,\partial_{j}\tilde{r}_{A}\bigg)\,.
\end{eqnarray}
By considering the metric (\ref{eq7}), the surface gravity takes the
following form
\begin{eqnarray}
\label{eq21}
\kappa=-\frac{1}{\tilde{r}_{A}}\bigg(1-\frac{\dot{\tilde{r}}_{A}}{2\,H\,\tilde{r}_{A}}\bigg)\,.
\end{eqnarray}
On the other hand, as it has been considered in Ref.~\cite{Awa14},
the general expression for the entropy in terms of area ($A$) is
given by
\begin{eqnarray}
\label{eq22} S=\frac{f(A)}{4G}\,.
\end{eqnarray}
Therefore, from equations (\ref{eq19}), (\ref{eq21}), and
(\ref{eq22}) we find
\begin{eqnarray}
\label{eq23}
T\,dS=-\frac{1}{2\,\pi\,\tilde{r}_{A}}\Bigg(1-\frac{\dot{\tilde{r}}_{A}}{2\,H\,\tilde{r}_{A}}\Bigg)d\Bigg(\frac{f(A)}{4\,G}\Bigg)
=-\frac{1}{2\,\pi\,\tilde{r}_{A}}\Bigg(1-\frac{\dot{\tilde{r}}_{A}}{2\,H\,\tilde{r}_{A}}\Bigg)\Bigg(\frac{d\,f(A)}{dA}\Bigg)
\Bigg(\frac{8\,\pi\,\tilde{r}_{A}}{4\,G}\,d\tilde{r}_{A}\Bigg)\,.\nonumber\\
\end{eqnarray}
From equation (\ref{eq9}), we have
$\dot{\tilde{r}}_{A}=-\tilde{r}_{A}^{2}\,\dot{H}$ and
$d\tilde{r}_{A}=-\tilde{r}_{A}^{2}\,\dot{H}\,dt$. By using these
equations and also equations (\ref{eq14}), (\ref{eq17}),
(\ref{eq18}) and (\ref{eq23}) we find the following expression for
the second Friedmann equation
\begin{eqnarray}
\label{eq24} \dot{H}\,\frac{d\,f(A)}{dA}=-4\pi\,G\,(\rho+p)\,.
\end{eqnarray}
From equation (\ref{eq11}) and integrating equation (\ref{eq24}) we
get
\begin{eqnarray}
\label{eq25} \rho=-\frac{3}{2\,G}\int \frac{f'(A)}{A^2}\,dA\,,
\end{eqnarray}
where $f'(A)=\frac{d\,f(A)}{dA}$. In the case with $f(A)=A$,
equations (\ref{eq24}) and (\ref{eq25}) give the standard Friedmann
equations as
\begin{eqnarray}
\label{eq26} \dot{H}=-4\pi\,G\,(\rho+p)\,,
\end{eqnarray}
\begin{eqnarray}
\label{eq27} H^{2}=\frac{8\pi\,G}{3}\rho\,.
\end{eqnarray}

Given that the presence of the minimal length shows itself in the
entropy-area relation, and then the entropy-area relation is used to
obtain the modified Friedmann equations, in the next section we
derive the modified Friedmann equation of the tachyon field in the
presence of the minimal length. As has been shown in
Ref.~\cite{Awa14}, this modification is obtained by considering a
GUP.

\section{The Friedmann Equations of the Tachyon Field in the Presence of the Minimal Length}

As we have mentioned in the Introduction, the presence of the
minimal length modifies the uncertainly principle as equation
(\ref{eq6}). This equation gives the minimal measurable length scale
as $\Delta x_{min}=2\sqrt{\beta}\,l_{Pl}$ and the following
expression for $\Delta p$~\cite{Awa14}
\begin{eqnarray}
\label{eq28} \Delta p \gtrsim
\frac{1}{2\beta\,l_{Pl}^{2}}\Bigg(\Delta x-\sqrt{(\Delta
x)^{2}-4\beta\,l_{Pl}^{2}}\Bigg)\,.
\end{eqnarray}
Also, the change in the area of a black hole that absorbs or emits a
quantum particle is given by $\Delta A\geq 8\pi\l_{Pl}^{2}\,E\,R$,
where $R$ is the particle's size and $E$ is its energy~\cite{Chr70}.
By considering the existence of a finite bound, implied by the
particle, we have $\Delta A_{min}\geq 8\pi\,l_{Pl}^{2}\,E\,\Delta
x$. Since the uncertainty principle $\Delta p\geq \frac{1}{\Delta
x}$ can be rewritten as $E\geq \frac{1}{\Delta
x}$~\cite{Med04,Ame04}, from equation (\ref{eq28}) we obtain
\begin{eqnarray}
\label{eq29} \Delta A_{min} \gtrsim \frac{8\pi\,\Delta
x}{2\beta}\Bigg(\Delta x-\sqrt{(\Delta
x)^{2}-4\beta\,l_{Pl}^{2}}\Bigg)\,.
\end{eqnarray}
Now, by assuming $\Delta x^{2}=\frac{A}{\pi}$, the change in the
area is given by~\cite{Awa14}
\begin{eqnarray}
\label{eq30} \Delta A_{min} \simeq \chi\,
\frac{8A}{2\beta}\Bigg(1-\sqrt{1-\frac{4\beta\,l_{Pl}^{2}\,\pi}{A}}\Bigg)\,.
\end{eqnarray}
By using the Bekenstein-Hawking entropy formula, the parameter
$\chi$ can be determined. Now we explore the relationship between
entropy and area. Given that the minimal increase of entropy is a
bit of information as $\Delta S_{min}=b=\ln (2)$~\cite{Ada04}, and
since from the Bekenstein-Hawking entropy formula we have
$\frac{b}{\chi}=2\,\pi$~\cite{Med04}, we get
\begin{eqnarray}
\label{eq31}\frac{dS}{dA}=\frac{\Delta S_{min}}{\Delta A_{min}}
=\frac{\pi}{\frac{2A}{\beta}\Bigg(1-\sqrt{1-\frac{4\beta\,l_{Pl}^{2}\,\pi}{A}}\Bigg)}\,.
\end{eqnarray}
From equation (\ref{eq22}), we find $f'(A)=4G\,\frac{dS}{dA}$.
Therefore, by using equation (\ref{eq31}) we can rewrite Friedmann
equations (\ref{eq24}) and (\ref{eq25}) as follows
\begin{eqnarray}
\label{eq32}\dot{H}\Bigg(\frac{\beta\,l_{Pl}^{2}}{2}\Bigg)\Bigg(\frac{H^{2}}{1-\sqrt{1-\beta\,l_{Pl}^{2}\,H^{2}}}\Bigg)=-4\pi\,G\,(\rho+p)\,,
\end{eqnarray}
\begin{eqnarray}
\label{eq33}\frac{H^{2}}{2}+\frac{1}{3\,\beta\,l_{Pl}^{2}}\,
\Bigg[1-\bigg(1-\beta\,l_{Pl}^{2}\,H^{2}\bigg)^{\frac{3}{2}}\Bigg]=\frac{8\pi\,G}{3}\,\rho\,,
\end{eqnarray}
where we have used $A=4\pi\,\tilde{r}_{A}^{2}$ (for more details
about derivation of these equations, see Ref.~\cite{Ada04}). Now, we
assume the energy component of the universe to be a tachyon field
$\phi$ with potential $V(\phi)$ and warp factor
$\lambda${~\cite{Li14,Kut00}}. In this case, the energy density and
pressure are
\begin{eqnarray}
\label{eq34}\rho=\frac{V(\phi)}{\sqrt{1-\lambda\dot{\phi}^{2}}}\,,
\quad p=-V(\phi)\,\sqrt{1-\lambda\dot{\phi}^{2}}\,,
\end{eqnarray}
leading to
\begin{eqnarray}
\label{eq35}\dot{H}\Bigg(\frac{\beta\,l_{Pl}^{2}}{2}\Bigg)\Bigg(\frac{H^{2}}{1-\sqrt{1-\beta\,l_{Pl}^{2}\,H^{2}}}\Bigg)=
-4\pi\,G\,\Bigg(\frac{\lambda\,V\,\dot{\phi}^{2}}{\sqrt{1-\lambda\dot{\phi}^{2}}}\Bigg)\,,
\end{eqnarray}
\begin{eqnarray}
\label{eq36}\frac{H^{2}}{2}+\frac{1}{3\,\beta\,l_{Pl}^{2}}\,
\Bigg[1-\bigg(1-\beta\,l_{Pl}^{2}\,H^{2}\bigg)^{\frac{3}{2}}\Bigg]
=\frac{8\pi\,G}{3}\,\Bigg(\frac{V}{\sqrt{1-\lambda\dot{\phi}^{2}}}\Bigg)\,.
\end{eqnarray}
Note that, if we use the Friedmann equations (\ref{eq35}) and
(\ref{eq36}) together with the conservation equation (\ref{eq11}),
we find that the effect of the minimal length doesn't show itself in
the equation of motion of the tachyon field up to the first order in
$l_{Pl}^{2}$. Therefore, we have
\begin{eqnarray}
\label{eq37}\frac{\lambda\,\ddot{\phi}}{1-\lambda\,\dot{\phi}^{2}}+3\,\lambda\,H\dot{\phi}
+\frac{V'(\phi)}{V(\phi)}=0\,,
\end{eqnarray}
as the equation of motion of the tachyon field. By these modified
Friedmann equations, the inflation parameters are modified too. In
the next section, we use these modified equations and study the
inflation in the presence of the minimal length.

\section{Tachyon Inflation in the Presence of the Minimal Length}

To solve some problems of the standard model of cosmology, a
primordial exponential phase of expansion in the early universe is
needed. In an exponential phase of expansion, the Hubble parameter
should be almost constant. This is because there was an end to the
inflation and after that, the universe became radiation-dominated.
In this way, some parameters for inflation have been defined which
are called the slow-roll parameters. These parameters are defined as
\begin{eqnarray}
\label{eq38}\epsilon=-\frac{\dot{H}}{H^{2}}\,,\quad
\eta=-\frac{\ddot{H}}{H\,\dot{H}}\,,
\end{eqnarray}
which in the inflation phase should satisfy the conditions
$|\epsilon,\,\eta|\ll 1$. In the single field inflation models, the
slow-roll conditions are corresponding to $\dot{\phi}^{2}\ll
V(\phi)$ and $\ddot{\phi}\ll 3H\dot{\phi}$. Under these conditions,
the Hubble parameter is almost constant ($H^{2}\simeq V(\phi)$) when
the potential is sufficiently flat. Now, the slow-roll parameters in
our model take the following forms
\begin{eqnarray}
\label{eq39}\epsilon\simeq\frac{\kappa^{2}\,V'^{2}}{18\,H^{4}\,\lambda\,V}\Bigg(1-\frac{\beta\,l_{Pl}^{2}}{16\,\pi}H^{2}
-\frac{\beta^{2}\,l_{Pl}^{4}}{256\pi^{2}}H^{4}\Bigg)^{-1}\,,
\end{eqnarray}

\begin{eqnarray}
\label{eq40}\eta=\frac{\Bigg[-\frac{\kappa^{2}}{27}\frac{V'^{2}\,V''}{2\dot{H}\,\lambda^{2}\,H^{4}\,V^{2}}
-\epsilon\,\bigg(1-\frac{3\beta\,l_{Pl}^{2}}{16\,\pi}H^{2}
-\frac{5\beta^{2}\,l_{Pl}^{4}}{256\pi^{2}}H^{4}\bigg)\Bigg]}{\Bigg[1-\frac{\beta\,l_{Pl}^{2}}{16\,\pi}H^{2}
-\frac{\beta^{2}\,l_{Pl}^{4}}{256\pi^{2}}H^{4}\Bigg]}\,,\nonumber\\
\end{eqnarray}
where we have used $\kappa^{2}=8\pi G$ and the small limit of
$\beta$ as $\beta\ll 4\pi H^{2}$. We can also use the well-known
e-folds number definition during the time interval between the
Hubble crossing of the physical scales ($hc$) and the end of
inflation ($end$) as
\begin{eqnarray}
\label{eq41}N=\int_{t_{hc}}^{t_{end}}
H\,dt=\int_{\phi_{hc}}^{\phi_{end}} \frac{H}{\dot{\phi}}\,d\phi\,,
\end{eqnarray}
where the parameter $H$ is given by equation (\ref{eq36}). Now, we
can study the perturbation parameters in this model to find more
information about the effects of the presence of the minimal length
in the theory. In the study of the perturbation parameters, the wave
number plays an important role. When we consider the minimal length
in the theory, the wave number gets modified. It has been shown
that, by considering the uncertainly relation (\ref{eq6}), the
modified commutation relation is given by~\cite{Kem95}
\begin{eqnarray}
\label{eq42}[X_{i},P_{j}]=i\Big(\delta_{ij}+\beta_{ijkl}\,l_{Pl}^{2}\,p^k\,p^l\Big)\,.
\end{eqnarray}
Now, we can write the position and momentum operators in the
position space as
\begin{eqnarray}
\label{eq43}X^i=x^{i}\,,\quad P^i=p^i(1+\beta\,l_{Pl}^{2}\, p^2)\,.
\end{eqnarray}
According to the equation (\ref{eq43}) and the relation $p=k$ (since we have set $\hbar=1$), we
find the modified wave number as~\cite{Ras23}
\begin{eqnarray}
\label{eq44}K^i=k^{i}(1+\beta\,l_{Pl}^{2}\, k^2)\,.
\end{eqnarray}
With this modified wave number, some perturbation parameters get
modified too. The first important parameter that we study here is
the scalar spectral index which is now defined as
\begin{eqnarray}
\label{eq45} n_{s}=1+\frac{d \ln {\cal{A}}_{s}}{d \ln
k(1+\beta\,l_{Pl}^{2}\, k^2)}\cong 1+\left(1+\beta\,l_{Pl}^{2}\,
k^2\right)\frac{d \ln {\cal{A}}_{s}}{d \ln
k}=1+\left(1+\beta\,l_{Pl}^{2}\, k^2\right)(-2\epsilon-\eta-s)\,,
\end{eqnarray}
where ${\cal{A}}_{s}$ is the amplitude of the scalar perturbations
defined as
\begin{eqnarray}
\label{eq46}
{\cal{A}}_{s}=\frac{H^{4}}{4\pi^{2}\,V\,(1-c_{s}^{2})}\,,
\end{eqnarray}
Note that, the effect of the minimal length in the amplitude is
encoded in $H$. The parameter $c_{s}$ in equation (\ref{eq46}) is
the sound speed given by
\begin{equation}
\label{eq47} c_s=\sqrt{1-\dot{\phi}^2}\,.
\end{equation}
Also, the parameter $s$ is defined as
\begin{equation}
\label{eq48} s=\frac{1}{H}\frac{\dot{c}_{s}}{c_{s}}\,,
\end{equation}
The tensor spectral index in the presence of the minimal length is
given by
\begin{eqnarray}
\label{eq49} n_{T}=\frac{d \ln {\cal{A}}_{T}}{d \ln
k(1+\beta\,l_{Pl}^{2}\, k^2)}\cong \left(1+\beta\,l_{Pl}^{2}\,
k^2\right)\frac{d \ln {\cal{A}}_{T}}{d \ln
k}=-\left(1+\beta\,l_{Pl}^{2}\, k^2\right)(2\epsilon)\,,
\end{eqnarray}
where ${\cal{A}}_{T}$ is the amplitude of the tensor perturbations
that is defined as
\begin{eqnarray}
\label{eq50} {\cal{A}}_{T}=\frac{2\kappa^{2}H^{2}}{\pi^{2}}\,.
\end{eqnarray}
Finally, the tensor-to-scalar ratio is given by
\begin{equation}
\label{eq51}
r=\frac{{\cal{A}}_{T}}{{\cal{A}}_{s}}=16\,c_{s}\,\epsilon\,=
-\frac{8c_{s}}{1+\beta l_{Pl}^{2}k^{2}}n_{T}
\end{equation}
This shows that the consistency relation gets modified in this
framework due to existence of a minimal measurable length. Now, we
are in the position that we can test our model's viability in
confrontation with recent observational data. We perform our
numerical analysis by choosing power-law, inverse power-law and
inverse exponential potentials.

\section{Observational Viability of the Model}

\begin{figure}[]
	\begin{center}
		\includegraphics[scale=0.12]{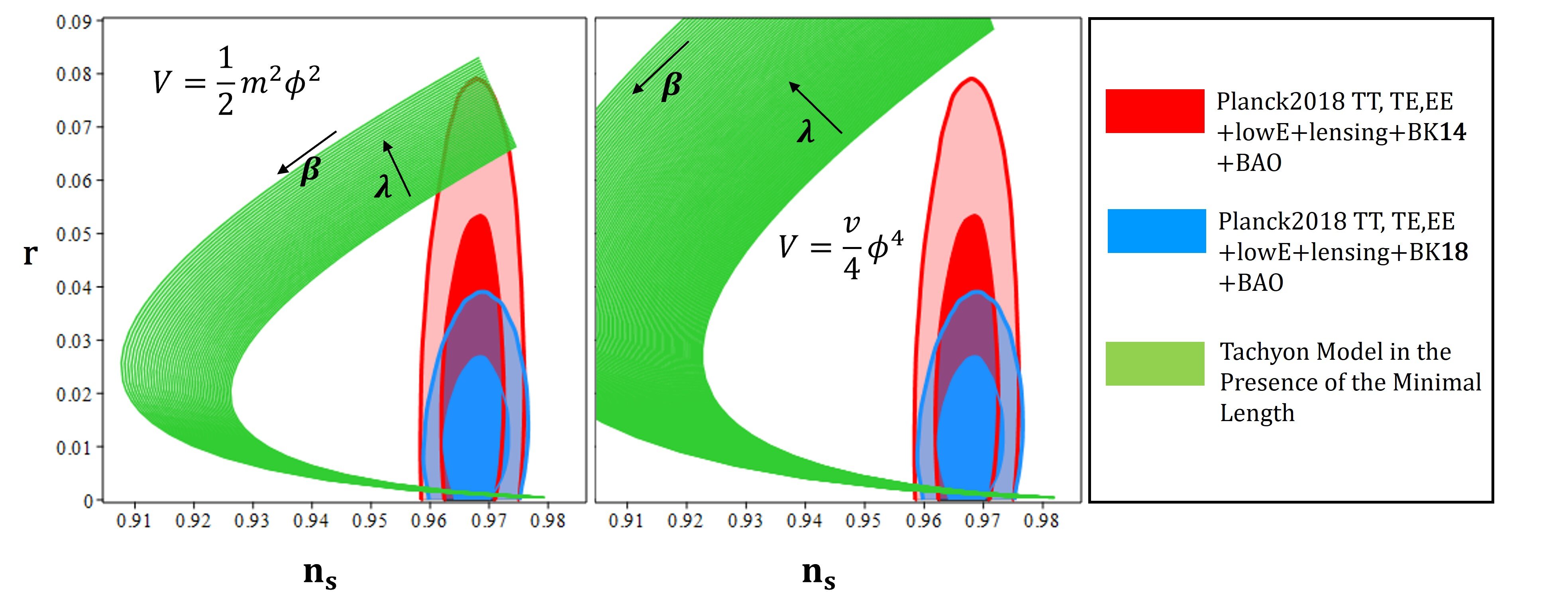}
	\end{center}
	\caption{\small {Tensor-to-scalar ratio versus the
			scalar spectral index for the tachyon inflation with $\phi^{2}$ and
			$\phi^{4}$ potentials in the presence of a minimal measurable
			length, in the background of Planck2018 TT, TE, EE
			+lowE+lensing+BK\textbf{14}+BAO~ and Planck2018 TT, TE, EE
			+lowE+lensing+BK\textbf{18}+BAO data.The arrows show
				the direction where the parameters increase. }}
	\label{fig1}
\end{figure}

When we consider a simple tachyon field with quadratic potential,
there is no consistency between the model and Planck2018 TT, TE,
EE+lowE+lensing+BAO+BK\textbf{14} observational data. For instance,
the value of the tensor-to-scalar ratio in the tachyon model with
$\phi^{2}$ and $\phi^{4}$ potentials for both $N=50$ and $N=60$ is
larger than the constraint released by Planck2018 TT, TE,
EE+lowE+lensing+BAO+BK\textbf{14} data. The observational upper
bound on the tensor-to-scalar ratio, based on
$\Lambda$CDM+$r$+$\frac{d n_{s}}{d\,\ln k}$ model, is
$r<0.072$~\cite{pl18a,pl18b}. Planck2018 TT, TE,
EE+lowE+lensing+BAO+BK\textbf{14} data implies the constraint on the
scalar spectral index as $n_{s}=0.9658\pm
0.0038$~\cite{pl18a,pl18b}. In this regard, the $r-n_{s}$ plot for
the tachyon model with $\phi^{2}$-potential lies beyond the
observational confidence levels and makes the model less favorable.
Also, Planck2018 TT, TE, EE+lowE+lensing+BAO+BK\textbf{18} data
implies a tighter constraint on the tensor-to-scalar ratio
($r<0.036$ at $95\%$ CL~\cite{Bi18a,Bi18b}) that rules out the
tachyon inflation with $\phi^{2}$ and $\phi^{4}$ potentials. Now
that we consider the minimal length in the tachyon model, by
considering the fact that it changes the slow-roll parameters and
therefore the perturbation parameters of the tachyon model, we
wonder if it is possible to find any consistency between the tachyon
model with quadratic potential and observational data. In this
regard, we assume $V=\frac{1}{2}\,m^{2}\,\phi^{2}$ and
$V=\frac{v}{4}\,\phi^{4}$ potentials, where $m=1.4\times 10^{13}$
GeV and $v=1.4\times 10^{-13}$~\cite{Man21}. By these potentials, we
find the slow-roll parameters (\ref{eq39}) and (\ref{eq40}) in terms
of the model's parameters. Then, we substitute these parameters in
the perturbations parameters $n_{s}$, $n_{T}$, and $r$. To perform
the numerical analysis, it is better to express the scalar field in
terms of other parameters. To this end, by using equations
(\ref{eq36}), (\ref{eq37}) and (\ref{eq41}), we find the scalar
field at horizon crossing in terms of $N$ and $\beta$. In this way,
we obtain the slow-roll parameters and therefore the perturbations
parameters in terms of $N$ and $\beta$. In this situation, it is
possible to test the observational viability of the model. Figure
\ref{fig1} shows $r-n_{s}$ behavior in the tachyon model with
$\phi^{2}$ and $\phi^{4}$ potentials, in the presence of the minimal
length. As the figure shows, the presence of the minimal length
makes the tachyon inflation model observationally viable. Of course,
there are some constraints on $\beta$ that lead to the consistency
of the model with observational data. To plot this figure, we have
adopted $N=60$ and $0.8<\lambda\leq 1$. Note that, in the full
numerical analysis, we have adopted $0<\lambda\leq 1$. However, in
plotting the figure, we consider a smaller range of $\lambda$, just
to show the $r-n_{s}$ behavior in confrontation with several data.
As this figure shows, $r-n_{s}$ plot in our model with $\phi^2$
potential, in two ranges of the parameter $\beta$ is consistent with
Planck2018 TT, TE, EE +lowE+lensing+BK\textbf{14} +BAO data. When we
consider Planck2018 TT, TE, EE +lowE+lensing+BK\textbf{18} +BAO
data, only one range of the model's parameters makes the model
consistent with observation. We have also performed the same
analysis with $\phi^4$ potential. In tables \ref{tab1}-\ref{tab4},
we have presented the observationally viable ranges of $\beta$, for
some sample values of $\lambda$ as $\lambda=0.1$, $0.4$, $0.7$ and
$1$ and for both of the mentioned potentials.

\begin{figure}[]
	\begin{center}
		\includegraphics[scale=0.12]{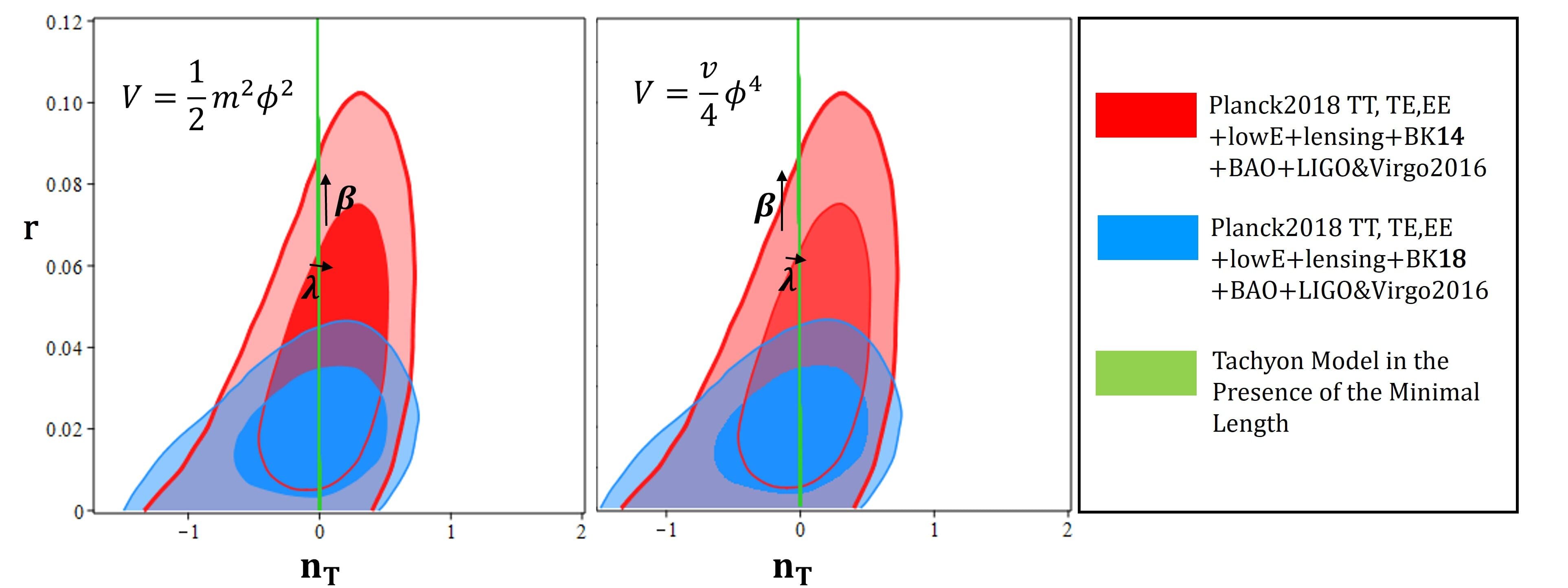}
	\end{center}
	\caption{\small {Tensor-to-scalar ratio versus the
			tensor spectral index for the tachyon inflation with $\phi^{2}$ and
			$\phi^{4}$ potentials in the presence of a minimal measurable
			length, in the background of Planck2018 TT, TE, EE
			+lowE+lensing+BK\textbf{14}+BAO+LIGO and Virgo2016 and Planck2018
			TT, TE, EE +lowE+lensing+BK\textbf{18}+BAO+LIGO and Virgo2016 data.
			The arrows show the direction where the
				parameters increase. }}
		\label{fig2}
	\end{figure}

\begin{figure}[]
	\begin{center}
		\includegraphics[scale=0.12]{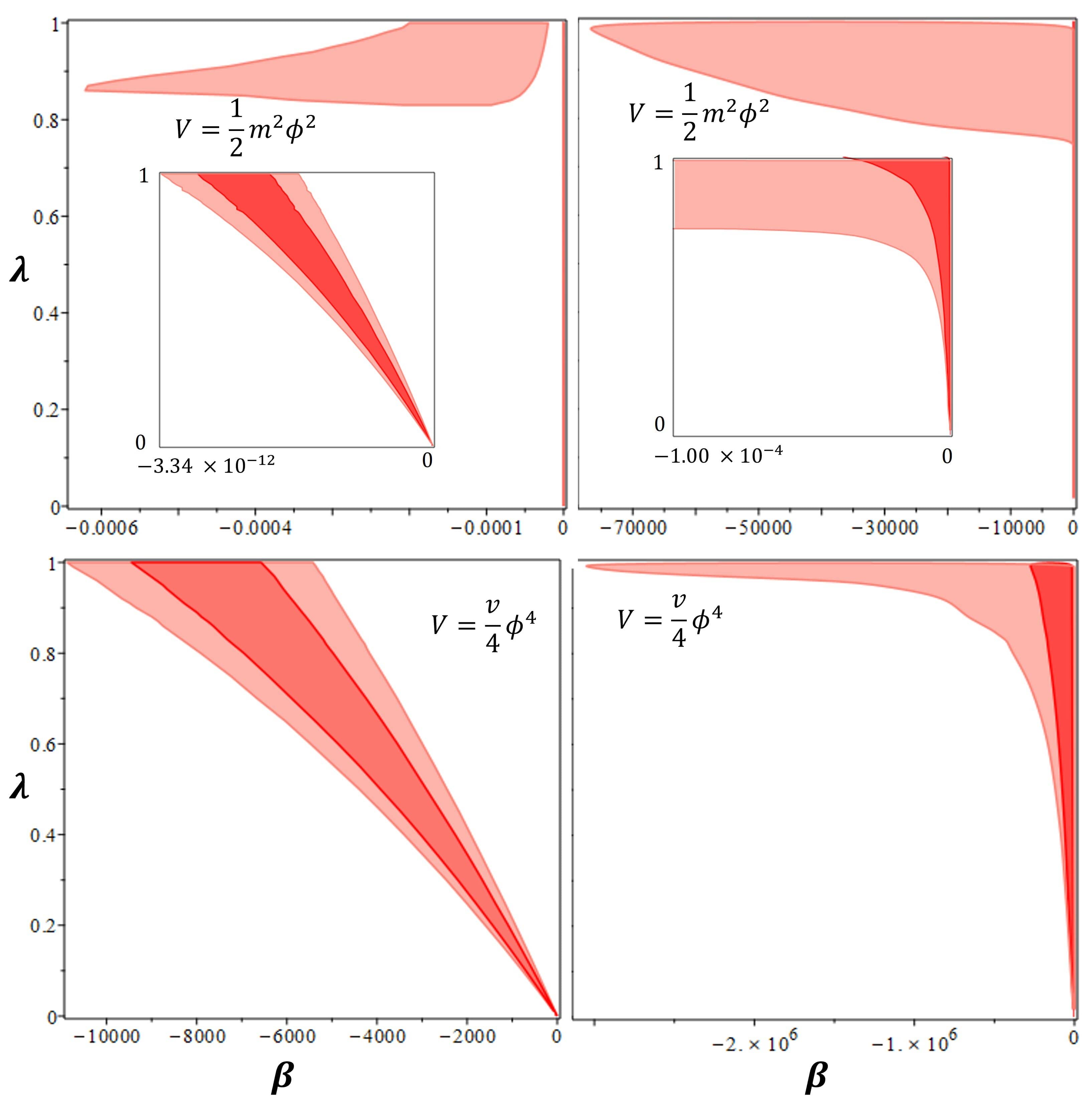}
	\end{center}
	\caption{\small {Left panels: ranges of the parameters
			$\lambda$ and $\beta$ for the tachyon inflation in the presence of a
			minimal length, leading to observationally viable values of
			$r-n_{s}$, in confrontation with Planck2018 TT, TE, and
			EE+lowE+lensing+BAO+BK\textbf{14} data at $68\%$ CL (dark red) and
			$95\%$ CL (light red). Right panels: ranges of the parameters
			$\lambda$ and $\beta$ for the tachyon inflation in the presence of
			the minimal length, leading to observationally viable values of
			$r-n_{T}$, in confrontation with Planck2018 TT, TE, and
			EE+lowE+lensing+BK\textbf{14}+BAO+LIGO and Virgo2016 data at $68\%$
			CL (dark red) and $95\%$ CL (light red). }}
		\label{fig3}
	\end{figure}

\begin{figure}[]
	\begin{center}
		\includegraphics[scale=0.12]{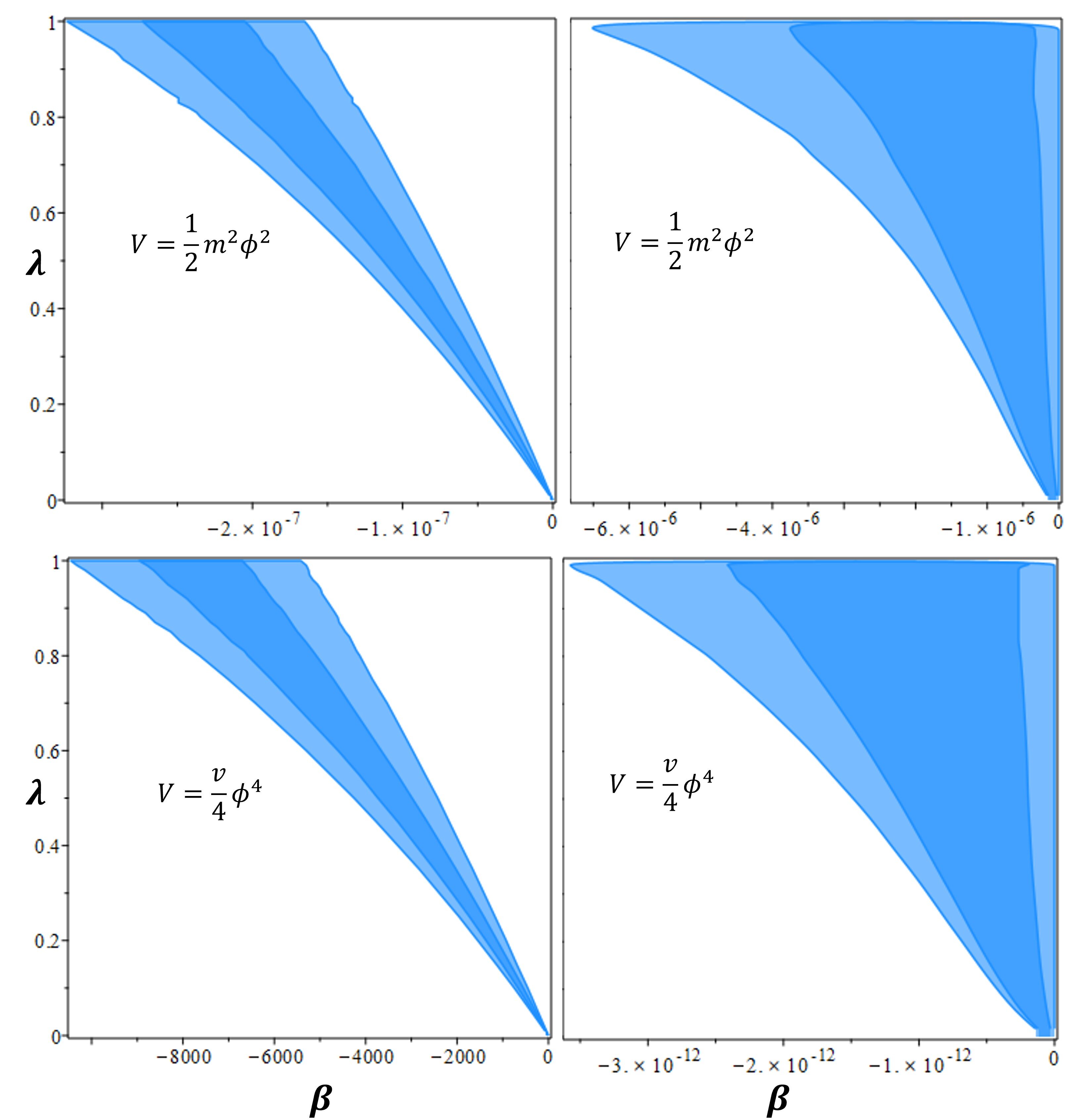}
	\end{center}
	\caption{\small {Left panels: ranges of the parameters
			$\lambda$ and $\beta$ for the tachyon inflation in the presence of a
			minimal length, leading to observationally viable values of
			$r-n_{s}$, in confrontation with Planck2018 TT, TE, and
			EE+lowE+lensing+BAO+BK\textbf{18} data at $68\%$ CL (dark blue) and
			$95\%$ CL (light blue). Right panels: ranges of the parameters
			$\lambda$ and $\beta$ for the tachyon inflation in the presence of
			the minimal length, leading to observationally viable values of
			$r-n_{T}$, in confrontation with Planck2018 TT, TE, and
			EE+lowE+lensing+BK\textbf{18}+BAO+LIGO and Virgo2016 data at $68\%$
			CL (dark blue) and $95\%$ CL (light blue).}}
	\label{fig4}
\end{figure}

The behavior of the tensor-to-scalar ratio versus the tensor
spectral index, in the background of both Planck2018 TT, TE, EE
+lowE+lensing+BK\textbf{14}+BAO +LIGO and Virgo2016 and Planck2018
TT, TE, EE +lowE+lensing+BK\textbf{18}+BAO +LIGO and Virgo2016 data
is shown in figure \ref{fig2}. As this figure shows, $r-n_{T}$ of
our model in some subspaces of the model's parameter space is
observationally viable. In this case also, the observationally
viable ranges of $\beta$, for some sample values of $\lambda$ are
shown in tables \ref{tab1}- \ref{tab4}. To demonstrate the
constraints on the model's parameters more clearly, in figures
\ref{fig3} and \ref{fig4} we have plotted $\lambda-\beta$ space in
confrontation with the observationally viable ranges of $r-n_{s}$
and $r-n_{T}$.

\begin{table*}[htbp]
\tiny\tiny\caption{\small{\label{tab1} Ranges of the parameter
$\beta$ in which the tensor-to-scalar ratio, scalar spectral index
and tensor spectral index of the tachyon model with $\phi^2$
potential in the presence of a minimal length are consistent with
different joint data sets.}}
\begin{center}
\tabcolsep=0.05cm\begin{tabular}{|c|c|c|c|c|c|} \hline \hline &&&&\\
& Planck2018 TT,TE,EE+lowE & Planck2018 TT,TE,EE+lowE & Planck2018
TT,TE,EE+lowE & Planck2018 TT,TE,EE+lowE
\\
& +lensing+BK\textbf{14}+BAO &
+lensing+BK\textbf{14}+BAO&lensing+BK\textbf{14}+BAO&lensing+BK\textbf{14}+BAO
\\
&  & &+LIGO$\&$Virgo2016 &LIGO$\&$Virgo2016
\\
\hline &&&&\\$\lambda$& $68\%$ CL & $95\%$ CL &$68\%$ CL & $95\%$ CL
\\
\hline\hline &&&&\\  $0.1$ &
$-2.10\times10^{-8}<\beta<-1.60\times10^{-8}$ &
$-2.33\times10^{-8}<\beta<-1.37\times10^{-8}
$ & $-6.60\times10^{-7}<\beta<-1.39\times10^{-9}$ & $-8.34\times10^{-7}<\beta<0$\\&&&& \\
\hline &&&&\\$0.4$& $-9.27\times10^{-8}<\beta<-6.95\times10^{-8}$ &
$-1.03\times10^{-7}<\beta<-5.80\times10^{-8}$
&$-2.34\times10^{-6}<\beta<-2.95\times10^{-9}$&
$-3.82\times10^{-6}<\beta<0$
\\ &&&& \\ \hline &&&&\\
$0.7$& $-1.79\times10^{-7}<\beta<-1.29\times10^{-7}$ &
$-2.03\times10^{-7}<\beta<-1.08\times10^{-7}
$ & $-6.26\times10^{-6}<\beta<-4.00\times10^{-9}$ & $-3.13\times10^{-5}<\beta<0$ \\ &&&& \\
\hline &&&&\\
$1$& $-2.90\times10^{-7}<\beta<-2.01\times10^{-7}$ &
$-3.35\times10^{-7}<\beta<-1.65\times10^{-7} $ &
$-5.21\times10^{-5}<\beta<-5.21\times10^{-9}$ &
$-7.82\times10^{4}<\beta<0$
\\
$$&  &
$-2.00\times10^{-4}<\beta<-2.00\times10^{-5} $ &  &  \\ &&&&\\
\hline  \hline
\end{tabular}
\end{center}
\end{table*}

\begin{table*}
\tiny\tiny\caption{\small{\label{tab2} Ranges of the parameter
$\beta$ in which the tensor-to-scalar ratio, scalar spectral index
and tensor spectral index of the tachyon model with $\phi^4$
potential in the presence of a minimal length are consistent with
different joint data sets.}}
\begin{center}
\tabcolsep=0.05cm\begin{tabular}{|c|c|c|c|c|c|} \hline \hline &&&&\\
& Planck2018 TT,TE,EE+lowE & Planck2018 TT,TE,EE+lowE&Planck2018
TT,TE,EE+lowE&Planck2018 TT,TE,EE+lowE
\\
& +lensing+BK\textbf{14}+BAO &
+lensing+BK\textbf{14}+BAO&lensing+BK\textbf{14}+BAO&lensing+BK\textbf{14}+BAO
\\
&  & &+LIGO$\&$Virgo2016 &LIGO$\&$Virgo2016
\\
\hline &&&&\\$\lambda$& $68\%$ CL & $95\%$ CL &$68\%$ CL & $95\%$ CL
\\
\hline\hline &&&&\\  $0.1$ &
$-6.97\times10^{2}<\beta<-5.33\times10^{2}$ &
$-7.72\times10^{2}<\beta<-4.56\times10^{2}
$ & $-1.84\times10^{4}<\beta<-4.18\times10^{2}$ & $-2.30\times10^{4}<\beta<0$\\ &&&& \\
\hline &&&&
\\$0.4$& $-3.03\times10^{3}<\beta<-2.26\times10^{3}$ & $-3.39\times10^{3}<\beta<-1.92\times10^{3}$
&$-5.85\times10^{4}<\beta<-9.06\times10^{2}$&
$-8.36\times10^{4}<\beta<0$
\\ &&&& \\ \hline &&&&\\
$0.7$& $-5.89\times10^{3}<\beta<-4.23\times10^{3}$ &
$-6.65\times10^{3}<\beta<-3.55\times10^{3}
$ & $-1.23\times10^{5}<\beta<-1.28\times10^{3}$ & $-2.43\times10^{5}<\beta<0$ \\ &&&&\\
\hline &&&&\\
$1$& $-9.44\times10^{3}<\beta<-6.57\times10^{3}$ &
$-1.08\times10^{4}<\beta<-5.42\times10^{3} $ &
$-2.82\times10^{5}<\beta<-1.49\times10^{3}$ &
$-3.11\times10^{6}<\beta<0$
\\ &&&& \\
\hline \hline
\end{tabular}
\end{center}
\end{table*}

\begin{table*}
\tiny\tiny\caption{\small{\label{tab3} Ranges of the parameter
$\beta$ in which the tensor-to-scalar ratio, scalar spectral index
and tensor spectral index of the tachyon model with $\phi^2$
potential in the presence of a minimal length are consistent with
different joint data sets.}}
\begin{center}
\tabcolsep=0.05cm\begin{tabular}{|c|c|c|c|c|c|} \hline \hline &&&&\\
& Planck2018 TT,TE,EE+lowE & Planck2018 TT,TE,EE+lowE&Planck2018
TT,TE,EE+lowE&Planck2018 TT,TE,EE+lowE
\\
& +lensing+BK\textbf{18}+BAO &
+lensing+BK\textbf{18}+BAO&lensing+BK\textbf{18}+BAO&lensing+BK\textbf{18}+BAO
\\
&  & &+LIGO$\&$Virgo2016 &LIGO$\&$Virgo2016
\\
\hline &&&&\\$\lambda$& $68\%$ CL & $95\%$ CL &$68\%$ CL & $95\%$ CL
\\
\hline\hline &&&&\\  $0.1$ &
$-2.00\times10^{-8}<\beta<-1.66\times10^{-8}$ &
$-2.26\times10^{-8}<\beta<-1.37\times10^{-8}
$ & $-4.40\times10^{-7}<\beta<-9.56\times10^{-9}$ & $-5.21\times10^{-7}<\beta<0$\\ &&&&\\
\hline &&&&\\$0.4$& $-8.74\times10^{-8}<\beta<-7.11\times10^{-8}$ &
$-9.98\times10^{-8}<\beta<-5.80\times10^{-8}$
&$-1.23\times10^{-6}<\beta<-1.98\times10^{-8}$&
$-1.61\times10^{-6}<\beta<0$
\\ &&&&\\ \hline &&&&\\
$0.7$& $-1.70\times10^{-7}<\beta<-1.31\times10^{-7}$ &
$-1.96\times10^{-7}<\beta<-1.07\times10^{-7}
$ & $-2.7\times10^{-6}<\beta<-2.60\times10^{-8}$ & $-3.28\times10^{-6}<\beta<0$ \\ &&&&\\
\hline &&&&\\
$1$& $-2.73\times10^{-7}<\beta<-2.05\times10^{-7}$ &
$-2.49\times10^{-7}<\beta<-1.65\times10^{-7} $ &
$-3.82\times10^{-6}<\beta<-3.13\times10^{-8}$ &
$-6.78\times10^{-6}<\beta<0$
 \\ &&&&\\
\hline \hline
\end{tabular}
\end{center}
\end{table*}

\begin{table*}
\tiny\tiny\caption{\small{\label{tab4} Ranges of the parameter
$\beta$ in which the tensor-to-scalar ratio, scalar spectral index
and tensor spectral index of the tachyon model with $\phi^4$
potential in the presence of a minimal length are consistent with
different joint data sets.}}
\begin{center}
\tabcolsep=0.05cm\begin{tabular}{|c|c|c|c|c|c|} \hline \hline &&&&\\
& Planck2018 TT,TE,EE+lowE & Planck2018 TT,TE,EE+lowE&Planck2018
TT,TE,EE+lowE&Planck2018 TT,TE,EE+lowE
\\
& +lensing+BK\textbf{18}+BAO &
+lensing+BK\textbf{18}+BAO&lensing+BK\textbf{18}+BAO&lensing+BK\textbf{18}+BAO
\\
&  & &+LIGO$\&$Virgo2016 &LIGO$\&$Virgo2016
\\
\hline &&&&\\$\lambda$& $68\%$ CL & $95\%$ CL &$68\%$ CL & $95\%$ CL
\\
\hline\hline &&&&\\  $0.1$ &
$-6.64\times10^{2}<\beta<-5.48\times10^{2}$ &
$-7.45\times10^{2}<\beta<-4.51\times10^{2}
$ & $-1.25\times10^{4}<\beta<-2.78\times10^{2}$ & $-1.48\times10^{4}<\beta<0$\\ &&&&\\
\hline &&&&\\$0.4$& $-2.89\times10^{3}<\beta<-2.33\times10^{3}$ &
$-3.29\times10^{3}<\beta<-1.91\times10^{3}$
&$-3.32\times10^{4}<\beta<-6.09\times10^{2}$&
$-4.18\times10^{4}<\beta<0$
\\ &&&&\\ \hline &&&&\\
$0.7$& $-5.57\times10^{3}<\beta<-4.35\times10^{3}$ &
$-6.41\times10^{3}<\beta<-3.53\times10^{3}
$ & $-5.59\times10^{4}<\beta<-7.84\times10^{2}$ & $-7.49\times10^{4}<\beta<0$ \\ &&&&\\
\hline &&&&\\
$1$& $-8.95\times10^{3}<\beta<-6.69\times10^{3}$ &
$-1.04\times10^{4}<\beta<-5.42\times10^{3} $ &
$-8.57\times10^{4}<\beta<-9.23\times10^{2}$ &
$-1.25\times10^{5}<\beta<0$
 \\ &&&&\\
\hline \hline
\end{tabular}
\end{center}
\end{table*}

\begin{figure}[]
	\begin{center}
		\includegraphics[scale=0.12]{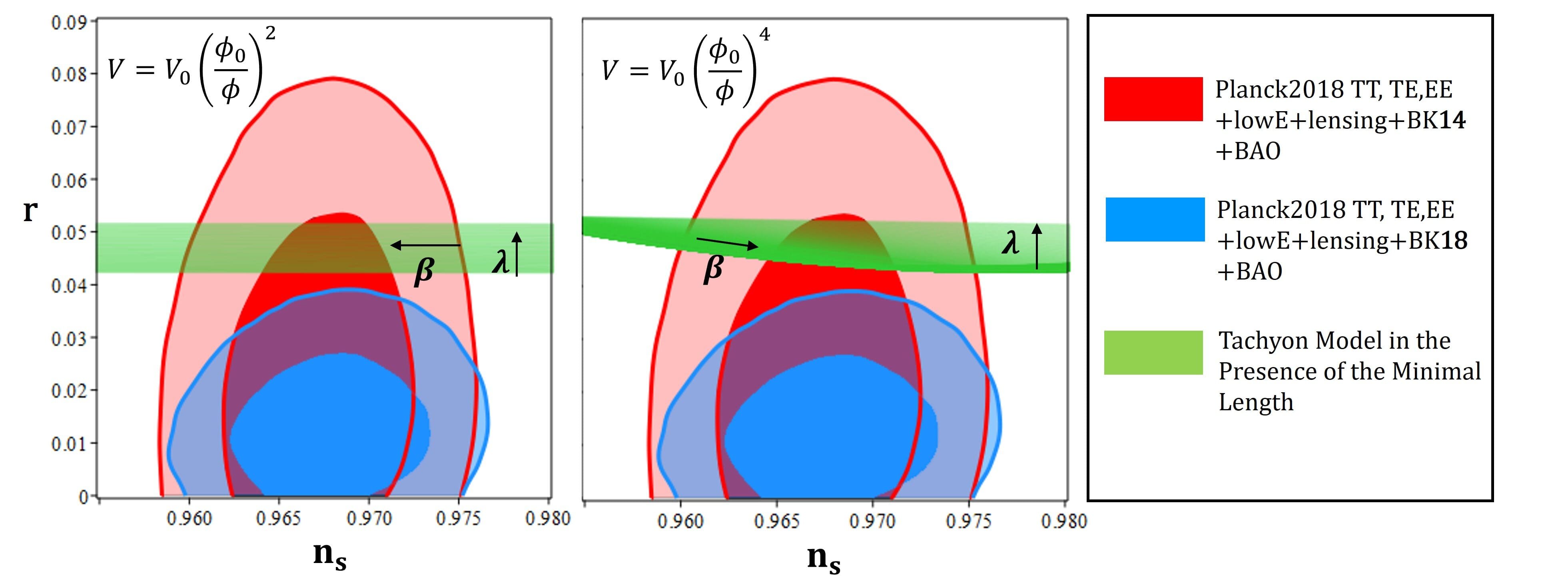}
	\end{center}
	\caption{\small {Tensor-to-scalar ratio versus the
			scalar spectral index for the tachyon inflation with $\phi^{-2}$ and
			$\phi^{-4}$ potentials in the presence of a minimal measurable
			length, in the background of Planck2018 TT, TE, EE
			+lowE+lensing+BK\textbf{14}+BAO~ and Planck2018 TT, TE, EE
			+lowE+lensing+BK\textbf{18}+BAO data. The arrows show the direction
			where the parameters increase.}}
	\label{fig5}
\end{figure}

\begin{figure}[]
	\begin{center}
		\includegraphics[scale=0.12]{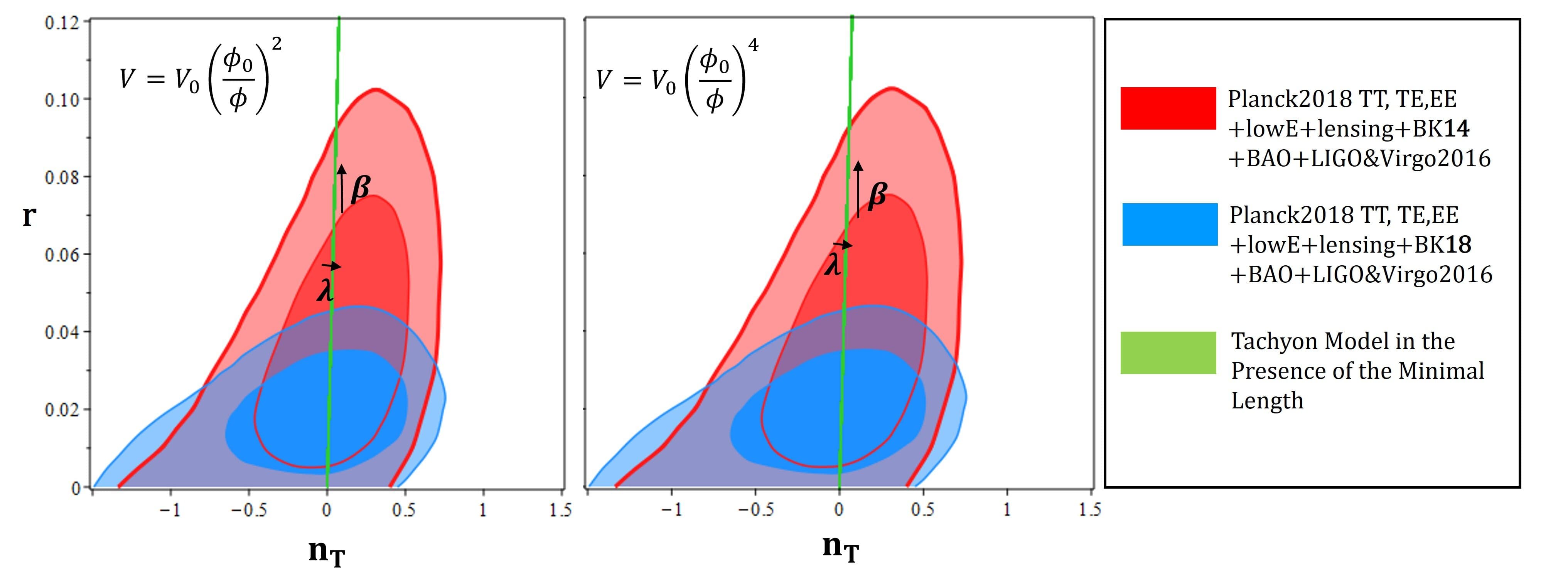}
	\end{center}
	\caption{\small {Tensor-to-scalar ratio versus the
			tensor spectral index for the tachyon inflation with $\phi^{-2}$ and
			$\phi^{-4}$ potentials in the presence of a minimal measurable
			length, in the background of Planck2018 TT, TE, EE
			+lowE+lensing+BK\textbf{14}+BAO+LIGO and Virgo2016 and Planck2018
			TT, TE, EE +lowE+lensing+BK\textbf{18}+BAO+LIGO and Virgo2016 data.
			The arrows show the direction where the parameters increase.}}
	\label{fig6}
\end{figure}

\begin{figure}[]
	\begin{center}
		\includegraphics[scale=0.12]{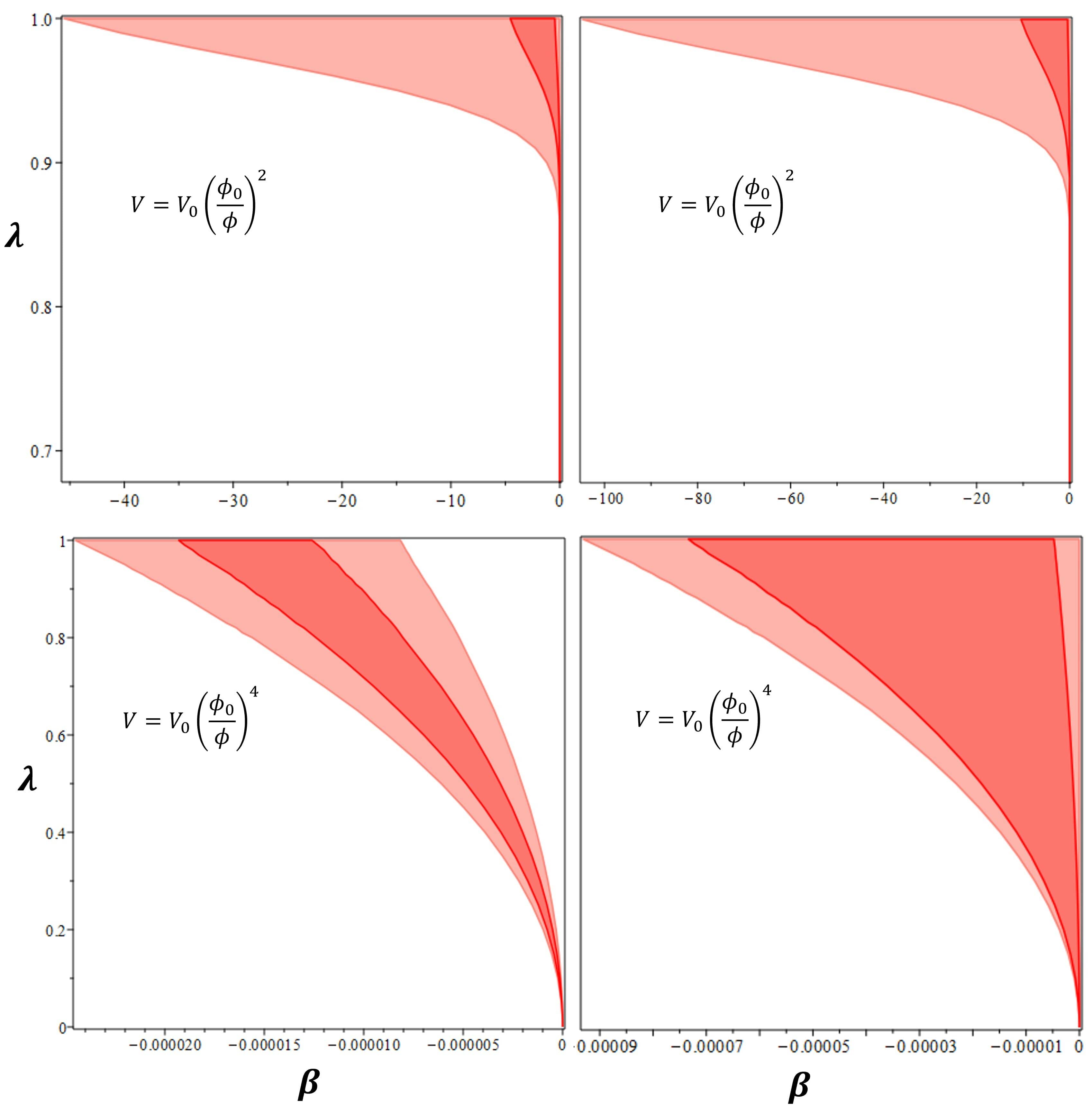}
	\end{center}
	\caption{\small {Left panels: ranges of the parameters
			$\lambda$ and $\beta$ for the tachyon inflation in the presence of a
			minimal length, leading to observationally viable values of
			$r-n_{s}$, in confrontation with Planck2018 TT, TE, and
			EE+lowE+lensing+BAO+BK\textbf{14} data at $68\%$ CL (dark red) and
			$95\%$ CL (light red). Right panels: ranges of the parameters
			$\lambda$ and $\beta$ for the tachyon inflation in the presence of
			the minimal length, leading to observationally viable values of
			$r-n_{T}$, in confrontation with Planck2018 TT, TE, and
			EE+lowE+lensing+BK\textbf{14}+BAO+LIGO and Virgo2016 data at $68\%$
			CL (dark red) and $95\%$ CL (light red).}}
	\label{fig7}
\end{figure}

\begin{figure}[]
	\begin{center}
		\includegraphics[scale=0.12]{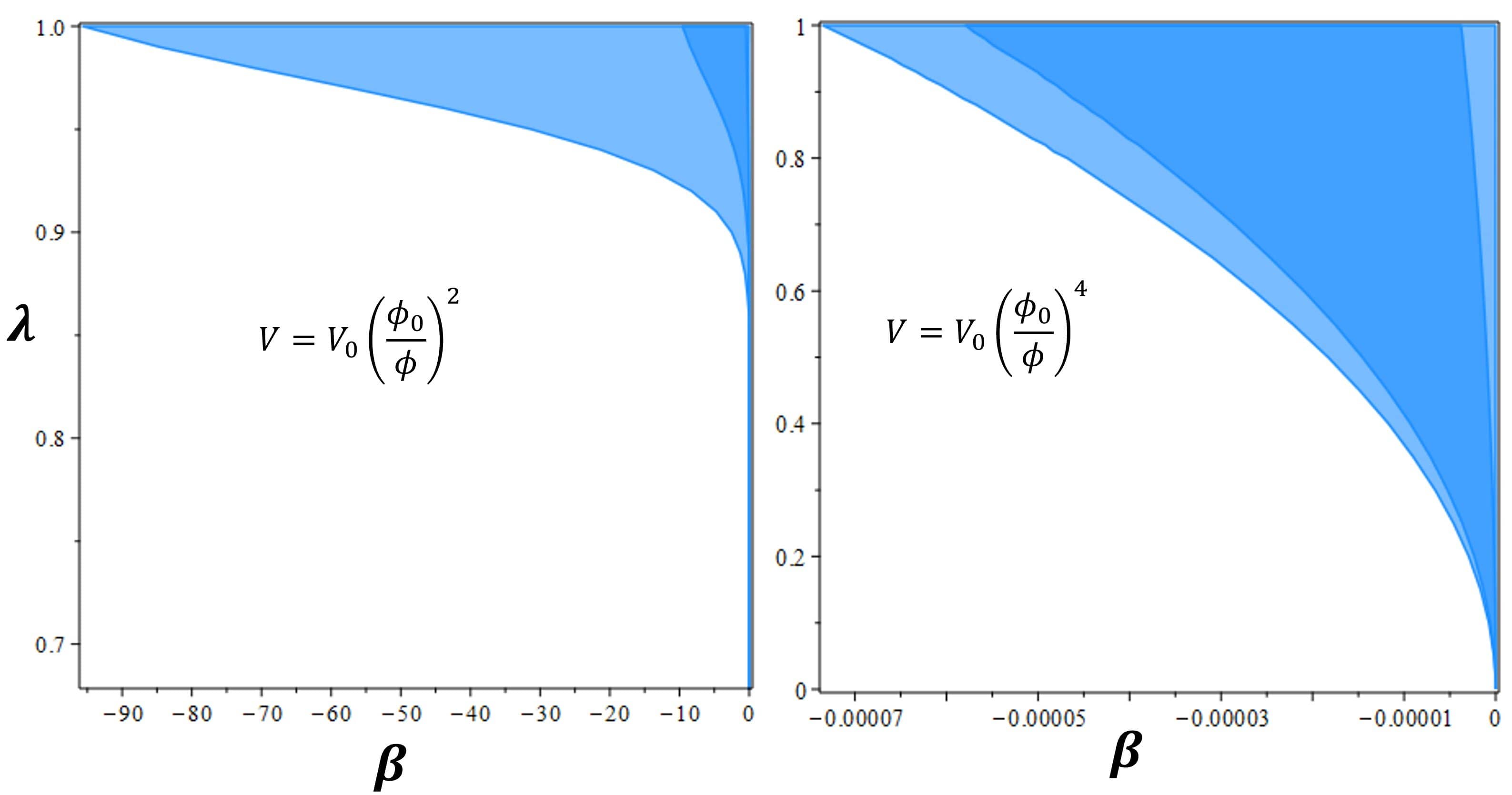}
	\end{center}
	\caption{\small {Ranges of the parameters $\lambda$ and
			$\beta$ for the tachyon inflation in the presence of a minimal
			length, leading to observationally viable values of $r-n_{T}$, in
			confrontation with Planck2018 TT, TE, and
			EE+lowE+lensing+BK\textbf{14}+BAO+LIGO and Virgo2016 data at $68\%$
			CL (dark blue) and $95\%$ CL (light blue).}}
	\label{fig8}
\end{figure}

{Other classes of potentials for the tachyon field, motivated by
string theory, are inverse power-law and inverse exponential
potentials. Tachyon inflation with these potentials has already been
studied in~\cite{Rez17}, where the authors have used Planck2015
datasets. By comparing their results with new observational data, we
find out that there is no consistency between the tachyon inflation
with these potentials and observational data. Now, we are going to
check if we can find any observational viability of the tachyon
inflation with inverse power-law and inverse exponential potentials
in the presence of the minimal length. In this regard, we first
consider $V=V_{0}\Big(\frac{\phi_{0}}{\phi}\Big)^{2}$ and
$V=V_{0}\Big(\frac{\phi_{0}}{\phi}\Big)^{4}$, where $V_{0}$ and
$\phi_{0}$ are some constants (re-scaled to 1). We follow the
structure for the $\phi^{2}$ and $\phi^{4}$ potentials and perform
some numerical analysis on our model. The behavior of $r-n_{s}$ for
the tachyon model with $\phi^{-2}$ and $\phi^{-4}$ potentials has
been shown in figure \ref{fig5}. As we can see from this figure, the
tachyon inflation with both considered potentials are only
consistent with Planck2018 TT, TE, EE +lowE+lensing+BK\textbf{14}
+BAO data at $68\%$ CL and $95\%$ CL. However, $r-n_{T}$ plots are
consistent with both Planck2018 TT, TE, EE
+lowE+lensing+BK\textbf{14} +BAO and Planck2018 TT, TE, EE
+lowE+lensing+BK\textbf{18} +BAO data (figure \ref{fig6}). For some
sample values of $\lambda$, we have obtained constraints on $\beta$,
summarized in tables \ref{tab5}-\ref{tab8}. Also, in figures
\ref{fig7} and \ref{fig8} we have plotted $\lambda-\beta$ space in
confrontation with the observationally viable ranges of $r-n_{s}$
and $r-n_{T}$. It should be noticed that, with $\phi^{-2}$
potential, the model has numerical results only for $\lambda>0.68$.}

\begin{table*}[htbp]
	{
\tiny\tiny\caption{\small{\label{tab5} Ranges of the parameter
$\beta$ in which the tensor-to-scalar ratio, scalar spectral index
and tensor spectral index of the tachyon model with $\phi^{-2}$
potential in the presence of a minimal length are consistent with
different joint data sets.}}
\begin{center}
\tabcolsep=0.05cm\begin{tabular}{|c|c|c|c|c|c|} \hline \hline &&&&\\
& Planck2018 TT,TE,EE+lowE & Planck2018 TT,TE,EE+lowE & Planck2018
TT,TE,EE+lowE & Planck2018 TT,TE,EE+lowE
\\
& +lensing+BK\textbf{14}+BAO &
+lensing+BK\textbf{14}+BAO&lensing+BK\textbf{14}+BAO&lensing+BK\textbf{14}+BAO
\\
&  & &+LIGO$\&$Virgo2016 &LIGO$\&$Virgo2016
\\
\hline &&&&\\$\lambda$& $68\%$ CL & $95\%$ CL &$68\%$ CL & $95\%$ CL
\\
\hline\hline &&&&\\  $0.7$ & $-1.05\times 10^{-1}<\beta<-1.04\times
10^{-1}$ & $-1.06\times 10^{-1}<\beta<-1.03 \times 10^{-1}
$ & $-1.96\times 10^{-1}<\beta<-1.44\times 10^{-1}$ & $-2.21\times 10^{-1}<\beta<0$\\&&&& \\
\hline &&&&\\$0.8$& $-1.07\times 10^{-1}<\beta<-1.05\times 10^{-1}$
& $-1.08\times 10^{-1}<\beta<-1.03\times 10^{-1}$ &$-2.40\times
10^{-1}<\beta<-1.07\times 10^{-1}$& $-2.48\times 10^{-1}<\beta<0$
\\ &&&& \\ \hline &&&&\\
$0.9$& $-2.31\times 10^{-1}<\beta<-1.10\times 10^{-1}$ &
$-1.56<\beta<-1.13\times 10^{-1}
$ & $-5.44\times 10^{-1}<\beta<-2.49\times 10^{-1}$ & $-3.30<\beta<0$ \\ &&&& \\
\hline &&&&\\
$1$& $-4.56<\beta<-5.53\times 10^{-1}$ & $-45.5<\beta<-1.14\times
10^{-1} $ & $-10.6<\beta<-0.601$ & $-104<\beta<0$
\\ &&&&\\
\hline  \hline
\end{tabular}
\end{center}}
\end{table*}

\begin{table*}[htbp]
	{
\tiny\tiny\caption{\small{\label{tab6} Ranges of the parameter
$\beta$ in which the tensor-to-scalar ratio, scalar spectral index
and tensor spectral index of the tachyon model with $\phi^{-4}$
potential in the presence of a minimal length are consistent with
different joint data sets.}}
\begin{center}
\tabcolsep=0.05cm\begin{tabular}{|c|c|c|c|c|c|} \hline \hline &&&&\\
& Planck2018 TT,TE,EE+lowE & Planck2018 TT,TE,EE+lowE & Planck2018
TT,TE,EE+lowE & Planck2018 TT,TE,EE+lowE
\\
& +lensing+BK\textbf{14}+BAO &
+lensing+BK\textbf{14}+BAO&lensing+BK\textbf{14}+BAO&lensing+BK\textbf{14}+BAO
\\
&  & &+LIGO$\&$Virgo2016 &LIGO$\&$Virgo2016
\\
\hline &&&&\\$\lambda$& $68\%$ CL & $95\%$ CL &$68\%$ CL & $95\%$ CL
\\
\hline\hline &&&&\\  $0.1$ & $-1.94\times 10^{-7}<\beta<-1.25\times
10^{-7}$ & $-2.45\times 10^{-7}<\beta<-8.10 \times 10^{-8}
$ & $-1.95\times 10^{-4}<\beta<-1.24\times 10^{-7}$ & $-2.47\times 10^{-6}<\beta<0$\\&&&& \\
\hline &&&&\\$0.4$& $-3.10\times 10^{-6}<\beta<-2.01\times 10^{-6}$
& $-3.90\times 10^{-6}<\beta<-1.31\times 10^{-6}$ &$-3.10\times
10^{-5}<\beta<-2.01\times 10^{-6}$& $-3.87\times 10^{-5}<\beta<0$
\\ &&&& \\ \hline &&&&\\
$0.7$& $-9.50\times 10^{-6}<\beta<-6.12\times 10^{-6}$ &
$-1.20\times 10^{-5}<\beta<-4.00\times 10^{-6}
$ & $-9.45\times 10^{-5}<\beta<-6.07\times 10^{-6}$ & $-1.16\times 10^{-4}<\beta<0$ \\ &&&& \\
\hline &&&&\\
$1$& $-1.93\times 10^{-5}<\beta<-1.26\times 10^{-5}$ & $-2.45\times
10^{-5}<\beta<-8.15\times 10^{-6} $ & $-1.90\times
10^{-4}<\beta<-1.26\times 10^{-5}$ & $-2.48\times 10^{-4}<\beta<0$
\\ &&&&\\
\hline  \hline
\end{tabular}
\end{center}}
\end{table*}

\begin{table*}[htbp]
	{
\tiny\tiny\caption{\small{\label{tab7} Ranges of the parameter
$\beta$ in which the tensor-to-scalar ratio, scalar spectral index
and tensor spectral index of the tachyon model with $\phi^{-2}$
potential in the presence of a minimal length are consistent with
different joint data sets.}}
\begin{center}
\tabcolsep=0.05cm\begin{tabular}{|c|c|c|c|c|c|} \hline \hline &&&&\\
& Planck2018 TT,TE,EE+lowE & Planck2018 TT,TE,EE+lowE & Planck2018
TT,TE,EE+lowE & Planck2018 TT,TE,EE+lowE
\\
& +lensing+BK\textbf{18}+BAO &
+lensing+BK\textbf{18}+BAO&lensing+BK\textbf{18}+BAO&lensing+BK\textbf{18}+BAO
\\
&  & &+LIGO$\&$Virgo2016 &LIGO$\&$Virgo2016
\\
\hline &&&&\\$\lambda$& $68\%$ CL & $95\%$ CL &$68\%$ CL & $95\%$ CL
\\
\hline\hline &&&&\\  $0.7$ & Not Consistent & Not Consistent & $-1.84\times 10^{-1}<\beta<-1.49\times 10^{-1}$ & $-2.09\times 10^{-1}<\beta<0$\\&&&& \\
\hline &&&&\\$0.8$& Not Consistent & Not Consistent  &$-2.16\times
10^{-1}<\beta<-1.11\times 10^{-1}$& $-2.37 \times 10^{-1}<\beta<0$
\\ &&&& \\ \hline &&&&\\
$0.9$& Not Consistent  &
Not Consistent  & $-4.72\times 10^{-1}<\beta<-2.57\times 10^{-1}$ & $-2.71<\beta<0$ \\ &&&& \\
\hline &&&&\\
$1$& Not Consistent  & Not Consistent  & $-9.75<\beta<-4.32\times
10^{-1}$ & $-95.6<\beta<0$
\\ &&&&\\
\hline  \hline
\end{tabular}
\end{center}}
\end{table*}

\begin{table*}[htbp]{
\tiny\tiny\caption{\small{\label{tab8} Ranges of the parameter
$\beta$ in which the tensor-to-scalar ratio, scalar spectral index
and tensor spectral index of the tachyon model with $\phi^{-4}$
potential in the presence of a minimal length are consistent with
different joint data sets.}}
\begin{center}
\tabcolsep=0.05cm\begin{tabular}{|c|c|c|c|c|c|} \hline \hline &&&&\\
& Planck2018 TT,TE,EE+lowE & Planck2018 TT,TE,EE+lowE & Planck2018
TT,TE,EE+lowE & Planck2018 TT,TE,EE+lowE
\\
& +lensing+BK\textbf{18}+BAO &
+lensing+BK\textbf{18}+BAO&lensing+BK\textbf{18}+BAO&lensing+BK\textbf{18}+BAO
\\
&  & &+LIGO$\&$Virgo2016 &LIGO$\&$Virgo2016
\\
\hline &&&&\\$\lambda$& $68\%$ CL & $95\%$ CL &$68\%$ CL & $95\%$ CL
\\
\hline\hline &&&&\\  $0.1$ &Not Consistent  & Not Consistent  & $-5.82\times 10^{-7}<\beta<-3.75\times 10^{-8}$ & $-7.35\times 10^{-7}<\beta<0$\\&&&& \\
\hline &&&&\\$0.4$& Not Consistent & Not Consistent  &$-9.30\times
10^{-6}<\beta<-6.03\times 10^{-7}$& $-1.17\times 10^{-5}<\beta<0$
\\ &&&& \\ \hline &&&&\\
$0.7$& Not Consistent  & Not Consistent  & $-2.85\times 10^{-5}<\beta<-1.83\times 10^{-6}$ & $-3.60\times 10^{-5}<\beta<0$ \\ &&&& \\
\hline &&&&\\
$1$& Not Consistent  & Not Consistent & $-5.79\times
10^{-5}<\beta<-3.78\times 10^{-6}$ & $-7.35\times 10^{-5}<\beta<0$
\\ &&&&\\
\hline  \hline
\end{tabular}
\end{center}}
\end{table*}

\begin{figure}[]
	\begin{center}
		\includegraphics[scale=0.12]{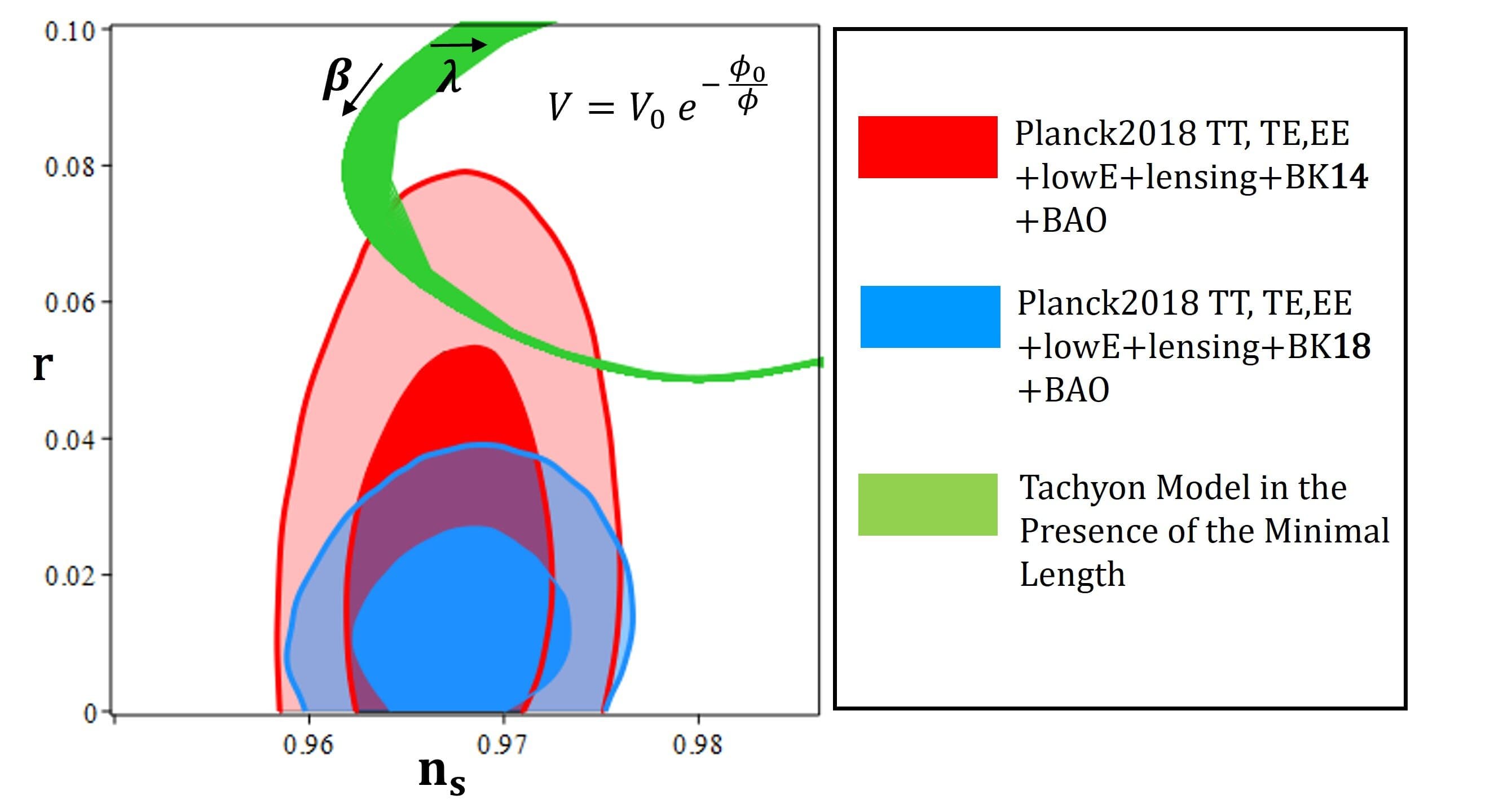}
	\end{center}
	\caption{\small {Tensor-to-scalar ratio versus the
			scalar spectral index for the tachyon inflation with inverse
			exponential potential in the presence of a minimal measurable
			length, in the background of Planck2018 TT, TE, EE
			+lowE+lensing+BK\textbf{14}+BAO~ and Planck2018 TT, TE, EE
			+lowE+lensing+BK\textbf{18}+BAO data. The arrows show the direction
			where the parameters increase.}}
	\label{fig9}
\end{figure}

\begin{figure}[]
	\begin{center}
		\includegraphics[scale=0.12]{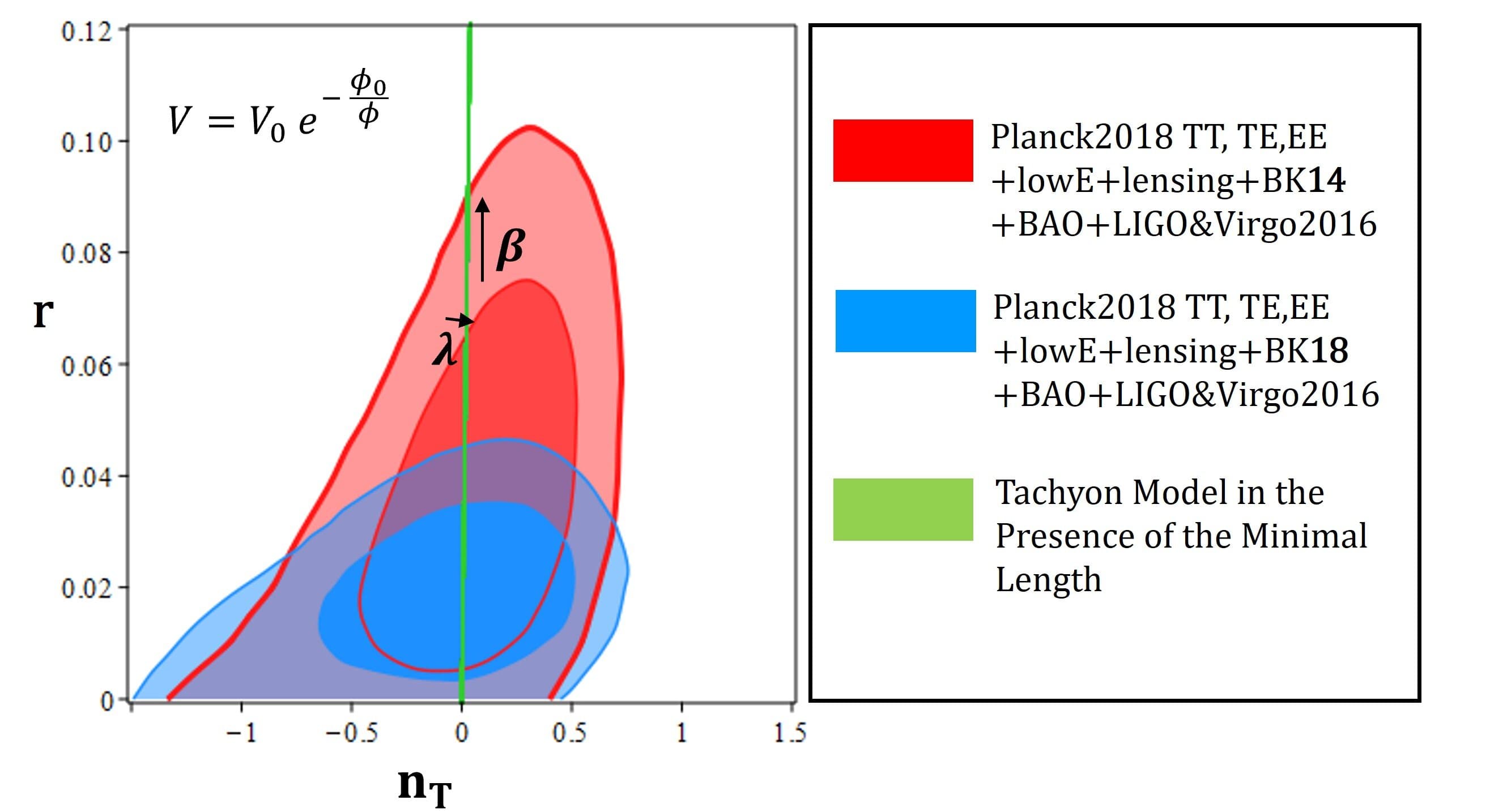}
	\end{center}
	\caption{\small {Tensor-to-scalar ratio versus the
			tensor spectral index for the tachyon inflation with inverse
			exponential potential in the presence of a minimal measurable
			length, in the background of Planck2018 TT, TE, EE
			+lowE+lensing+BK\textbf{14}+BAO+LIGO and Virgo2016 and Planck2018
			TT, TE, EE +lowE+lensing+BK\textbf{18}+BAO+LIGO and Virgo2016 data.
			The arrows show the direction where the parameters increase.}}
	\label{fig10}
\end{figure}

\begin{figure}[]
	\begin{center}
		\includegraphics[scale=0.12]{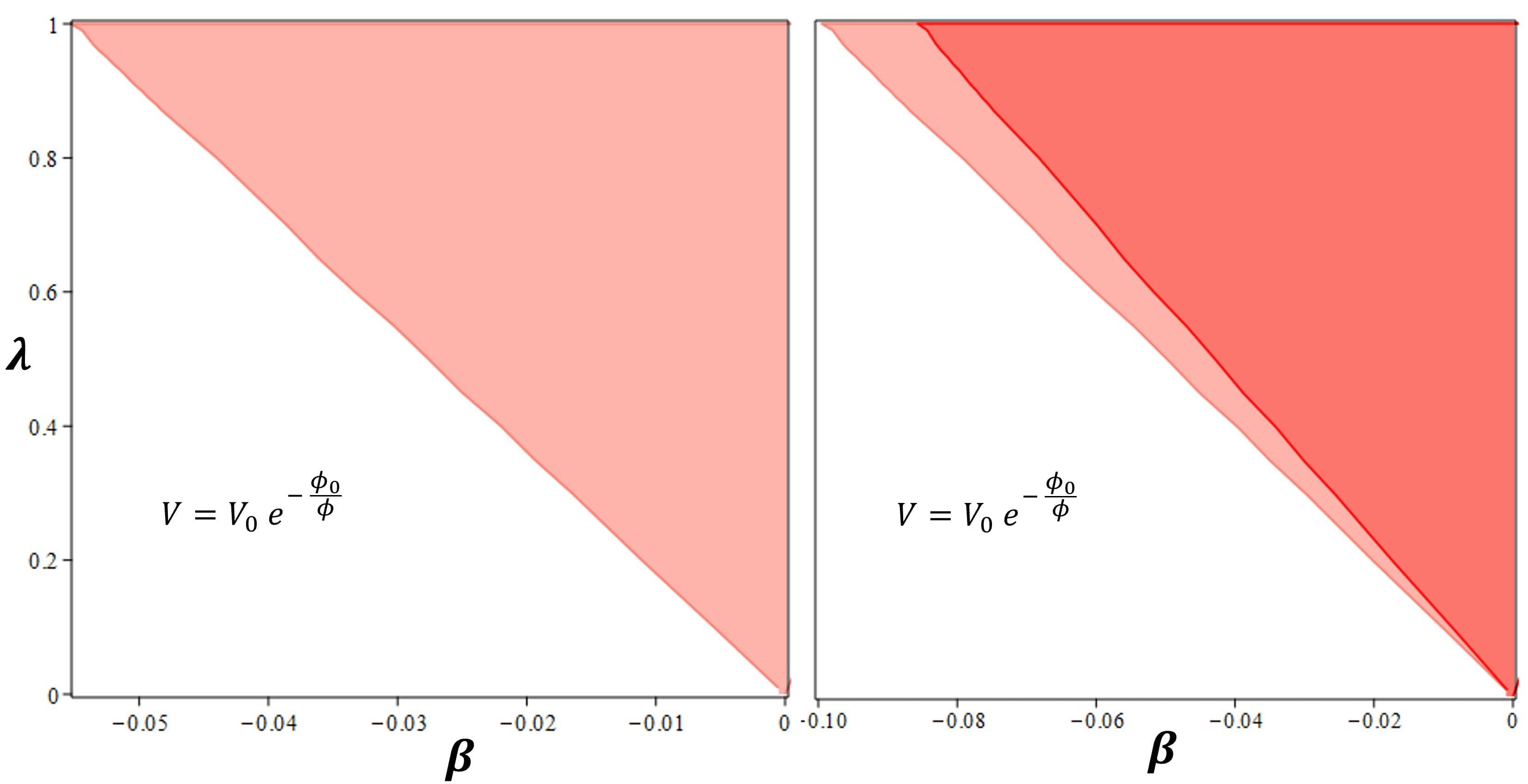}
	\end{center}
	\caption{\small {Left panels: ranges of the parameters
			$\lambda$ and $\beta$ for the tachyon inflation in the presence of a
			minimal length, leading to observationally viable values of
			$r-n_{s}$, in confrontation with Planck2018 TT, TE, and
			EE+lowE+lensing+BAO+BK\textbf{14} data at $95\%$ CL (light red).
			Right panels: ranges of the parameters $\lambda$ and $\beta$ for the
			tachyon inflation in the presence of the minimal length, leading to
			observationally viable values of $r-n_{T}$, in confrontation with
			Planck2018 TT, TE, and EE+lowE+lensing+BK\textbf{14}+BAO+LIGO and
			Virgo2016 data at $68\%$ CL (dark red) and $95\%$ CL (light red).}}
	\label{fig11}
\end{figure}

\begin{figure}[]
	\begin{center}
		\includegraphics[scale=0.12]{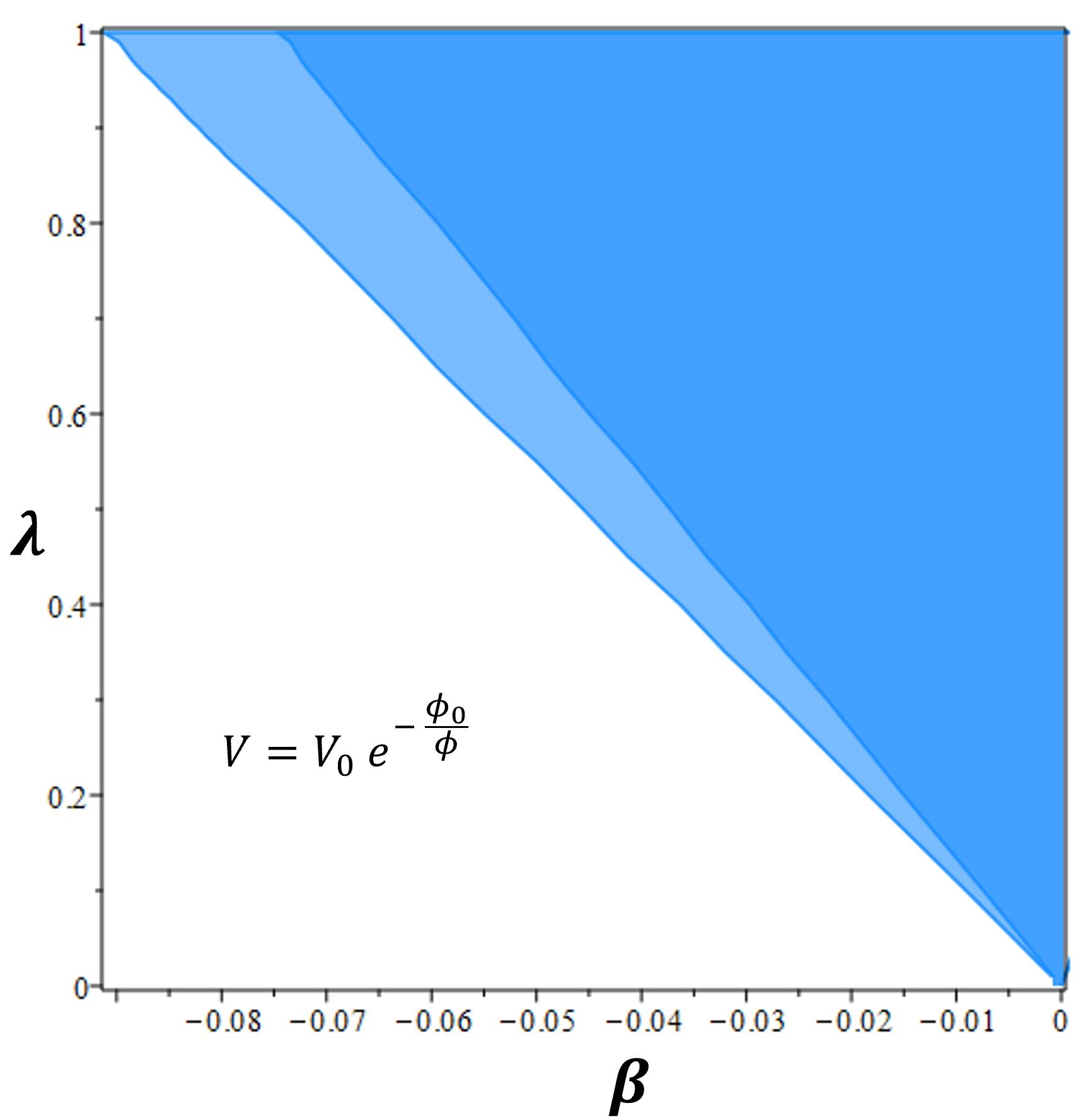}
	\end{center}
	\caption{\small {Ranges of the parameters $\lambda$ and
			$\beta$ for the tachyon inflation in the presence of a minimal
			length, leading to observationally viable values of $r-n_{T}$, in
			confrontation with Planck2018 TT, TE, and
			EE+lowE+lensing+BK\textbf{14}+BAO+LIGO and Virgo2016 data at $68\%$
			CL (dark blue) and $95\%$ CL (light blue).}}
	\label{fig12}
\end{figure}

{Now, we consider the inverse exponential potential as
$V=V_{0}\exp(-\frac{\phi}{\phi_{0}})$, where again $V_{0}$ and
$\phi_{0}$ are some constants that we re-scale them to 1. Similar to
the previous potentials, we perform numerical analysis on the
$r-n_{s}$ and $r-n_{T}$. The results are shown in figures \ref{fig9}
and \ref{fig10}. As we see from these figures, in this case also,
$r-n_{s}$ is only consistent with Planck2018 TT, TE, EE
+lowE+lensing+BK\textbf{14}+BAO data at $95\%$ CL. However,
$r-n_{T}$ is consistent with both Planck2018 TT, TE, EE
+lowE+lensing+BK\textbf{14}+BAO+LIGO and Virgo2016 and Planck2018
TT, TE, EE +lowE+lensing+BK\textbf{18}+BAO+LIGO and Virgo2016 data.
As before, we obtained some constraints on $\beta$ for some sample
values of $\lambda$, summarized in tables \ref{tab9} and
\ref{tab10}. Figures \ref{fig11} and \ref{fig12} demonstrate
$\lambda-\beta$ space in confrontation with the observationally
viable ranges of $r-n_{s}$ and $r-n_{T}$ for the model with inverse
exponential potential.}

\begin{table*}[htbp]
	{
\tiny\tiny\caption{\small{\label{tab9} Ranges of the parameter
$\beta$ in which the tensor-to-scalar ratio, scalar spectral index
and tensor spectral index of the tachyon model with
$\exp\left(-\frac{\phi}{\phi_{0}}\right)$ potential in the presence
of a minimal length are consistent with different joint data sets.}}
\begin{center}
\tabcolsep=0.05cm\begin{tabular}{|c|c|c|c|c|c|} \hline \hline &&&&\\
& Planck2018 TT,TE,EE+lowE & Planck2018 TT,TE,EE+lowE & Planck2018
TT,TE,EE+lowE & Planck2018 TT,TE,EE+lowE
\\
& +lensing+BK\textbf{14}+BAO &
+lensing+BK\textbf{14}+BAO&lensing+BK\textbf{14}+BAO&lensing+BK\textbf{14}+BAO
\\
&  & &+LIGO$\&$Virgo2016 &LIGO$\&$Virgo2016
\\
\hline &&&&\\$\lambda$& $68\%$ CL & $95\%$ CL &$68\%$ CL & $95\%$ CL
\\
\hline\hline &&&&\\  $0.1$ & Not Consistent  & $-5.50\times
10^{-3}<\beta<-2.20 \times 10^{-8}
$ & $-8.25\times 10^{-3}<\beta<-1.74\times 10^{-8}$ & $-9.90\times 10^{-3}<\beta<0$\\&&&& \\
\hline &&&&\\$0.4$& Not Consistent  & $-2.20\times
10^{-2}<\beta<-9.10\times 10^{-8}$ &$-3.41\times
10^{-2}<\beta<-7.46\times 10^{-8}$& $-3.96\times 10^{-2}<\beta<0$
\\ &&&& \\ \hline &&&&\\
$0.7$& Not Consistent  & $-3.86\times 10^{-2}<\beta<-1.59\times
10^{-7}
$ & $-5.98\times 10^{-2}<\beta<-1.34\times 10^{-7}$ & $-6.94\times 10^{-2}<\beta<0$ \\ &&&& \\
\hline &&&&\\
$1$& Not Consistent  & $-5.53\times 10^{-2}<\beta<-2.30\times
10^{-7} $ & $-8.57\times 10^{-2}<\beta<-1.02\times 10^{-7}$ &
$-9.95\times 10^{-2}<\beta<0$
\\ &&&&\\
\hline  \hline
\end{tabular}
\end{center}}
\end{table*}

\begin{table*}[htbp]
	{
\tiny\tiny\caption{\small{\label{tab10} Ranges of the parameter
$\beta$ in which the tensor-to-scalar ratio, scalar spectral index
and tensor spectral index of the tachyon model with
$\exp\left(-\frac{\phi}{\phi_{0}}\right)$ potential in the presence
of a minimal length are consistent with different joint data sets.}}
\begin{center}
\tabcolsep=0.05cm\begin{tabular}{|c|c|c|c|c|c|} \hline \hline &&&&\\
& Planck2018 TT,TE,EE+lowE & Planck2018 TT,TE,EE+lowE & Planck2018
TT,TE,EE+lowE & Planck2018 TT,TE,EE+lowE
\\
& +lensing+BK\textbf{18}+BAO &
+lensing+BK\textbf{18}+BAO&lensing+BK\textbf{18}+BAO&lensing+BK\textbf{18}+BAO
\\
&  & &+LIGO$\&$Virgo2016 &LIGO$\&$Virgo2016
\\
\hline &&&&\\$\lambda$& $68\%$ CL & $95\%$ CL &$68\%$ CL & $95\%$ CL
\\
\hline\hline &&&&\\  $0.1$ & Not Consistent  & Not Consistent & $-7.42\times 10^{-3}<\beta<-2.97\times 10^{-8}$ & $-9.07\times 10^{-3}<\beta<0$\\&&&& \\
\hline &&&&\\$0.4$& Not Consistent & Not Consistent  &$-2.97\times
10^{-2}<\beta<-1.22\times 10^{-7}$& $-3.63 \times 10^{-2}<\beta<0$
\\ &&&& \\ \hline &&&&\\
$0.7$& Not Consistent  &
Not Consistent  & $-5.21\times 10^{-2}<\beta<-1.57\times 10^{-7}$ & $-6.36\times 10^{-2}<\beta<0$ \\ &&&& \\
\hline &&&&\\
$1$& Not Consistent  & Not Consistent & $-7.46\times
10^{-2}<\beta<-1.32\times 10^{-7}$ & $-9.12\times 10^{-2}<\beta<0$
\\ &&&&\\
\hline  \hline
\end{tabular}
\end{center}}
\end{table*}

{Note that, the warp factor $\lambda$ is not
enough itself to make the tachyon model observationally viable. In
fact, the authors of Ref.~\cite{Li14} have obtained the
observationally viable domain in parameter space of $\lambda-m^{2}$
and also $\lambda-v$. However, for values of $\lambda$ and $v$ that
we have fixed here, the tachyon model with $0<\lambda\leq 1$ is not
consistent with Planck2018 data. This fact has also been confirmed
with our analysis. In our all numerical analysis and for all
$0<\lambda\leq 1$ we obtain $\beta\neq 0$ to get observational
viability.}

According to our numerical analysis, the tachyon model in the
presence of the minimal length is consistent with several data sets
as follows:
\begin{itemize}

\item{ The model with $\phi^2$ potential is consistent with Planck2018 TT, TE, EE +lowE+lensing+BK\textbf{14} +BAO data at
$95\%$ CL, if $-3.35\times 10^{-7}\leq \beta \leq -1.37\times
10^{-8}$ and $-2.00\times 10^{-4}\leq\beta\leq -2.00\times 10^{-5}$,
depending on the value of $\lambda$.}

\item{ This model is consistent with the same data at $68\%$ CL if
$-2.90\times 10^{-7}\leq\beta\leq -1.60\times 10^{-8}$.}

\item{ The model with $\phi^4$ potential is consistent with Planck2018 TT, TE, EE +lowE+lensing+BK\textbf{14} +BAO data at
$95\%$ CL, if $-1.08\times 10^{4}\leq \beta \leq -4.56\times
10^{2}$.}

\item{ With $\phi^4$ potential, there is consistency with the data at
$68\%$ CL, if $-9.44\times 10^{3}\leq \beta \leq -5.53\times
10^{2}$.}

\item{With $\phi^2$ potential, if we consider Planck2018 TT, TE, EE +lowE+lensing+BK\textbf{18} +BAO data,
the constraint on the parameter $\beta$ is $-2.49\times 10^{-7}\leq
\beta\leq -1.37\times 10^{-8}$ at $95\%$ CL.}

\item{With the same data, the constraint on $\beta$ is $-2.73\times 10^{-7}\leq\beta\leq -1.66\times 10^{-8}$ at $68\%$
CL, depending on the value of $\lambda$.}

\item{With $\phi^4$ potential, if we consider Planck2018 TT, TE, EE +lowE+lensing+BK\textbf{18} +BAO data,
the constraint on the parameter $\beta$ is $-1.04\times 10^{4}\leq
\beta\leq -4.51\times 10^{2}$ at $95\%$ CL.}

\item{With the same data, the constraint on $\beta$ is $-8.95\times 10^{3}\leq\beta\leq -4.51\times 10^{2}$ at $68\%$
CL, depending on the value of $\lambda$.}

\item{{With $\phi^{-2}$ potential and based on Planck2018 TT, TE, EE +lowE+lensing+BK\textbf{14} +BAO data,
the model is observationally viable if $-45.5\leq\beta\leq
-1.03\times 10^{-1}$ at $95\%$ CL.}}

\item{{With the same data and same potential, we find observational viability if $-4.56\leq\beta\leq
-1.04\times 10^{-1}$ at $68\%$ CL.}}

\item{{Considering $\phi^{-4}$ potential and Planck2018 TT, TE, EE +lowE+lensing+BK\textbf{14} +BAO data,
our setup is consistent with observation if $-2.45\times
10^{-5}\leq\beta\leq -8.10\times 10^{-8}$ at $95\%$ CL.}}

\item{{With the same potential and data, there is consistency at
$68\%$ CL if $-1.93\times 10^{-5}\leq\beta\leq -1.25\times
10^{-7}$.}}

\item{{The tachyon model with inverse exponential potential is consistence with Planck2018 TT, TE, EE +lowE+lensing+BK\textbf{14} +BAO data
at $95\%$ CL if $-5.53\times 10^{-2}\leq\beta\leq -2.20\times
10^{-8}$ at $95\%$ CL.}}

\item{{If we consider Planck2018 TT, TE, EE +lowE+lensing+BK\textbf{18} +BAO data,
our model with $\phi^{-2}$, $\phi^{-4}$ and inverse exponential
potentials is not observationally viable.}}

{In the above list, we have only considered
Planck2018 TT, TE, EE +lowE+lensing+BK\textbf{14}(\textbf{18}) +BAO
data. This is because the mentioned constraints overlap with
Planck2018 TT, TE, and
EE+lowE+lensing+BK\textbf{14}(\textbf{18})+BAO+LIGO and Virgo2016
data. Therefore, the list actually covers the constraints from both
data.} In conclusion, considering the effects
of the minimal length in the tachyon inflation makes the model
observationally viable, at least in some subspaces of the model
parameters space. { Based on our analysis and the obtained
constraints, it seems that when we consider $\phi^{4}$ potential,
the modification to the uncertainty relation needed to make the
model viable is almost big and of the order of $10^{2}-10^{3}$.
However, with $\phi^{2}$, $\phi^{-2}$ and $\phi^{-4}$ potentials,
this modification is very small and of the order of
$10^{-8}-10^{-7}$. Another case is corresponding to the inverse
exponential potential leading to the modification of the order of
$10^{-8}-10^{-3}$ which is small. Therefore, it seems that the order
of the modification to the uncertainty is model-dependent.}

\end{itemize}

\section{Conclusion}

In this paper, we have studied tachyon inflation by considering the
quantum gravitational effects, encoded in the existence of a minimal
measurable length, that seem to be important in the early universe.
To this end, we have started by reviewing the construction of the
modified Friedmann equations in the context of the first law of
thermodynamics. In this reconstruction, the modified Friedmann
equations are obtained by considering the work density in terms of
the trace of the energy-momentum tensor and also assuming a general
expression for the entropy. On the other hand, when we deal with the
early universe, the quantum gravitational effects are naturally
important. In this regard, and by considering a minimal measurable
length in the theory, we have presented the modified Friedmann
equations of the tachyon model. The reason for dealing with tachyon
inflation is the string nature of tachyon which obviously has
something to do with quantum gravitational effects. Especially, the
minimal length is of the order of string length. The presence of the
minimal length shows itself by a deviation parameter $\beta$, in the
Friedmann equations. When the Friedmann equations are modified, as a
result, the slow-roll parameters of inflation are modified too. We
have obtained the modified slow-roll parameters in the tachyon model
where the effect of the presence of minimal length appears
explicitly. Since the existence of a minimal length modifies the
momentum operator, and therefore the wave number, the perturbation
parameters get modified. We have obtained the scalar and tensor
spectral indices and also the tensor-to-scalar ratio for the tachyon
model in the presence of the minimal length. After that, we have
performed some numerical analysis on these parameters. In this
regard, we have considered several types of potential as $\phi^2$ and
$\phi^4$, $\phi^{-2}$, $\phi^{-4}$, and inverse exponential, and performed our analysis based on these potentials. We
have compared $r-n_{s}$ behavior of our model with both Planck2018
TT, TE, EE +lowE+lensing+BK\textbf{14} +BAO and Planck2018 TT, TE,
EE +lowE+lensing+BK\textbf{18} +BAO data. We have also analyzed
$r-n_{T}$ behavior in confrontation with Planck2018 TT, TE, EE
+lowE+lensing+BK\textbf{14}+BAO +LIGO and Virgo2016 and Planck2018
TT, TE, EE +lowE+lensing+BK\textbf{18}+BAO +LIGO and Virgo2016 data.
In this way, we have obtained the observationally viable ranges of
the model's parameters $\lambda$ and $\beta$ in comparison to
several data sets. Based on our analysis, it is seen that by
considering the presence of the minimal length in the tachyon
inflation makes the model observationally viable. This is an
important result since tachyon inflation for the mentioned potentials was ruled out in the absence of minimal
length.

{\textbf{ACKNOWLEDGMENTS}\\
We thank the referees for the very insightful comments that have
improved the quality of the paper considerably.}


\begin{thebibliography}{100}
	
\bibitem{Ama89} D. Amati, M. Ciafaloni \& G. Veneziano, Phys. Lett. B \textbf{216}, 41-47 (1989).

{\bibitem{Cas15} R. Casadio, O. Micu \& P. Nicolini, Fundam. Theor. Phys. \textbf{178}, 293-322 (2015).}

{\bibitem{Gro88} D. J. Gross \& P. F. Mende, Nucl. Phys. B \textbf{303}, 407-454 (1988).}

{\bibitem{Yo89} T. Yoneya, Mod. Phys. Lett. A \textbf{4}, 1587 (1989).}

{\bibitem{Ko90} K. Konishi, G. Paffuti \& P. Provero, Phys. Lett. B \textbf{234}, 276-284 (1990).}

{\bibitem{Rov95} C. Rovelli \& L. Smolin, Nucl. Phys. B \textbf{442}, 593-622 (1995).}

{\bibitem{Ash97} A. Ashtekar \& J. Lewandowski, Class. Quant. Grav. \textbf{14}, A55-A82 (1997).}

{\bibitem{Ash98} A. Ashtekar \& J. Lewandowski, Adv. Theor. Math. Phys. \textbf{1}, 388-429 (1998)}. 

{\bibitem{Mo09} L. Modesto, Class. Quant. Grav. \textbf{26}, 242002 (2009).} 

{\bibitem{Lau05} O. Lauscher \& M. Reuter, JHEP \textbf{10}, 050 (2005).} 

{\bibitem{Re06} M. Reuter \& J. M. Schwindt, JHEP \textbf{01}, 070 (2006).}

{\bibitem{Per10} R. Percacci \& G. P. Vacca, Class. Quant. Grav. \textbf{27}, 245026 (2010).}

{\bibitem{Eic19} A. Eichhorn, Front. Astron. Space Sci. \textbf{5}, 47 (2019).}

{\bibitem{Sei99} N. Seiberg \& E. Witten, JHEP \textbf{09}, 032 (1999).}

{\bibitem{Con99} A. Connes, J. Math. Phys. \textbf{41}, 3832-3866 (2000).}

{\bibitem{Ash03} A. Ashtekar, S. Fairhurst \& J. L. Willis, Class. Quant. Grav. \textbf{20}, 1031-1062 (2003).}

{\bibitem{Hos13} S. Hossenfelder, Living Rev. Rel. \textbf{16}, 2 (2013).}

{\bibitem{Kof02} L. Kofman \& A. Linde, JHEP \textbf{0207}, 004 (2002).}

{\bibitem{Kem95} A. Kempf, G. Mangano \& R. B. Mann, Phys. Rev. D \textbf{52}, 1108-1118 (1995).}

{\bibitem{No12} K. Nozari and A. Etemadi, Phys. Rev. D \textbf{85}, 104029 (2012).}

{\bibitem{Ame02} G. Amelino-Camelia, Int. J. Mod. Phys. D \textbf{11}, 35-60 (2002).}
 
{\bibitem{Cor05} J. L. Cortes and J. Gamboa, Phys. Rev. D \textbf{71}, 065015 (2005). }


\bibitem{Jac95} T. Jacobson, Phys. Rev. Lett. \textbf{75}, 1260 (1995).

\bibitem{Ver00} E. Verlinde, [arXiv:hep-th/0008140] (2000).

\bibitem{Cai03} R. G. Cai \& Y. S. Myung, Phys. Rev. D \textbf{67}, 124021 (2003).

\bibitem{Noj01} S. Nojiri, S. D. Odintsov \& S. Ogushi, Int. J. Mod. Phys. A \textbf{16}, 5085 (2001).

\bibitem{Noj02} S. Nojiri, S. D. Odintsov \& S. Ogushi, Int. J. Mod. Phys. A \textbf{17}, 4809 (2002).

\bibitem{Cve02} M. Cvetic, S. Nojiri \& S.D. Odintsov, Nucl. Phys. B \textbf{628}, 295 (2002).

\bibitem{Cai05} R. G. Cai \& S. P. Kim, JHEP \textbf{0502}, 050 (2005).

\bibitem{Med04} A. J. M. Medved \& E. C. Vagenas, Phys. Rev. D \textbf{70}, 124021 (2004).

\bibitem{Adl01} R. J. Adler, P. Chen \& D. I. Santiago, Gen. Rel. Grav. \textbf{33}, 2101 (2001).

\bibitem{Cav03} M. Cavaglia, S. Das \& R. Maartens, Class. Quant. Grav. \textbf{20}, L205 (2003).

\bibitem{Cav04} M. Cavaglia \& S. Das, Class. Quant. Grav. \textbf{21}, 4511 (2004).

\bibitem{Maj11} B. Majumder, Phys. Lett. B \textbf{703}, 402 (2011).

\bibitem{Ali12a} A. F. Ali, JHEP \textbf{1209}, 067 (2012).

\bibitem{Ali12b} A. F. Ali, H. Nafie \& M. Shalaby, Europhys. Lett. \textbf{100}, 20004 (2012).

\bibitem{Cam04} G. Amelino-Camelia, M. Arzano \& A. Procaccini, Phys. Rev. D \textbf{70}, 107501 (2004).

\bibitem{Awa14} A. Awad \& A. Farag Ali, JHEP \textbf{1406}, 093 (2014).

\bibitem{Gut81} A. Guth, Phys. Rev. D \textbf{23}, 347 (1981).

\bibitem{Lin82} A. D. Linde, Phys. Lett. B \textbf{108}, 389 (1982).

\bibitem{Alb82} A. Albrecht \& P. Steinhard, Phys. Rev. D \textbf{48}, 1220 (1982).

\bibitem{Lin90} A. D. Linde, \emph{Particle Physics and Inflationary Cosmology} (Harwood Academic Publishers, Chur, Switzerland) (1990).

\bibitem{Lid00a} A. Liddle \& D. Lyth, \emph{Cosmological Inflation and Large-Scale Structure}, (Cambridge University Press) (2000).

\bibitem{Lid97} J. E. Lidsey, A. R. Liddle, E. W. Kolb, E. J. Copeland, T. Barreiro \& M. Abney, Rev. Mod. Phys. \textbf{69}, 373 (1997).

\bibitem{Rio02} A. Riotto, [arXiv:hep-ph/0210162] (2002).

\bibitem{Lyt09} D. H. Lyth \& A. R. Liddle, \emph{The Primordial Density Perturbation} (Cambridge University Press) (2009).

\bibitem{Mal03} J. M. Maldacena, JHEP \textbf{0305}, 013 (2003).

\bibitem{Sen99} A. Sen, JHEP \textbf{10}, 008 (1999).

\bibitem{Sen02a} A. Sen, JHEP \textbf{07}, 065 (2002).

\bibitem{Sen02b} A. Sen, Modern Physics Letters A \textbf{17}, 1797 (2002).

\bibitem{Gib02} G. W. Gibbons, Phys. Lett. B \textbf{537}, 1 (2002).

\bibitem{Noj03} S. Nojiri \& S. D. Odintsov, Phys. Lett. B \textbf{571}, 1 (2003).

\bibitem{Bou16} Z. Bouabdallaoui, A. Errahmani, M. Bouhmadi-L$\acute{o}$pez \& T. Ouali, Phys. Rev. D \textbf{94}, 123508 (2016).

\bibitem{Rez17} K. Rezazadeh, K. Karami \& S. Hashemi, Phys. Rev. D \textbf{95}, 103506 (2017).

{\bibitem{Maj04} M. Majumdar \& A.-C. Davis, Phys. Rev. D \textbf{69}, 103504 (2004).}

{\bibitem{Li14} S. Li \& A. R. Liddle, JCAP \textbf{03}, 044 (2014).}

{\bibitem{Ade14} P. A. R. Ade, et al., A \& A \textbf{571}, A22 (2014).}

{\bibitem{Ade16} P. A. R. Ade, et al., A \& A \textbf{594}, A20 (2016).}

\bibitem{Duf97} M. J. Duff, NuPhB \textbf{125}, 334 (1977).

\bibitem{pl18a} N. Aghanim, Y. Akrami, M. Ashdown, et al., A \& A \textbf{641}, A6 (2020).

\bibitem{pl18b} Y. Akrami, F. Arroja, M. Ashdown, et al., A \& A \textbf{641}, A10 (2020).

\bibitem{Bi18a} P.A.R. Ade, Z. Ahmed, M. Amiri, D. Barkats, R. B. Thakur, et al. Phys. Rev. Lett. \textbf{127}, 151301 (2021).

\bibitem{Bi18b} D. Paoletti, F. Finelli, J. Valiviita \& M. Hazumi, Phys. Rev. D \textbf{106}, 083528 (2022).

\bibitem{Cha10} S. Chakraborty, R. Biswas \& N. Mazumder, Nuovo Cim. B \textbf{125}, 1209 (2011).

\bibitem{Cha11} S. Chakraborty, N. Mazumder \& R. Biswas, Gen. Rel. Grav. \textbf{43}, 1827 (2011).

\bibitem{Hay98} S. A. Hayward Class. Quant. Grav. \textbf{15}, 3147 (1998).

\bibitem{Bak00} D. Bak, S.-J. Rey, Class. Quant. Grav. \textbf{17}, 83 (2000).

\bibitem{Chr70} D. Christodoulou, Phys. Lett. \textbf{25}, 1596 (1970).

\bibitem{Ame04} G. Amelino-Camelia, M. Arzano \& A. Procaccini, Phys. Rev. D \textbf{70}, 107501 (2004).

\bibitem{Ada04} C. Adami, [arXiv:quant-ph/0405005] (2004).

{\bibitem{Kut00} D. Kutasov, M. Marino \& G. Moore, JHEP \textbf{0010}, 045 (2000).}

\bibitem{Ras23} N. Rashidi, M. Roushan \& K. Nozari, EPL \textbf{142}, 39001 (2023).

\bibitem{Man21} A. Mantziris, T. Markkanen \& A. Rajantie, Journal of Cosmology and Astroparticle Physics \textbf{03}, 077 (2021).



\end{thebibliography}
\end{document}